\documentclass[useAMS,usenatbib]{mn2e}
\usepackage{epsfig,natbib}
\usepackage{amssymb,amsmath}
\usepackage{natbib}
\usepackage{graphicx}
\usepackage{color}
\usepackage{rotating}

\usepackage{fixltx2e}

\DeclareMathOperator\erf{erf}

\title[An algorithm to build mock galaxy catalogues]
{An algorithm to build mock galaxy catalogues using MICE simulations}
\author[J.~Carretero et al.]
{\parbox[t]{\textwidth}{J. Carretero$^{1,2}$\thanks{Email: jcarrete@ieec.uab.es}, F. J. Castander$^1$,  E. Gazta\~naga$^1$, M. Crocce$^1$ \& P. Fosalba$^1$}\vspace*{6pt}\\
$^1$Institut de Ci\`{e}ncies de l'Espai, IEEC-CSIC, Campus UAB, F. de Ci\`{e}ncies, Torre C5 par-2, Barcelona 08193, Spain\\
$^2$Port d{'}Informaci\'{o} Cient\'{i}fica (PIC) Edifici D, Universitat Aut\`{o}noma de Barcelona (UAB), E-08193 Bellaterra (Barcelona), Spain.}

\date{\today}
\pagerange{000-000}
\pubyear{0000}
\begin{document}

\maketitle
\label{firstpage}

\begin{abstract}

We present a method to build mock galaxy catalogues starting from a halo catalogue that uses halo occupation distribution (HOD) recipes as well as the subhalo abundance matching (SHAM) technique. Combining both prescriptions we are able to push the absolute magnitude of the resulting catalogue to fainter luminosities than using just the SHAM technique and can interpret our results in terms of the HOD modelling. We optimize the method by populating with galaxies friends-of-friends dark matter haloes extracted from the Marenostrum Institut de Ci\`{e}ncies de l'Espai (MICE) dark matter simulations and comparing them to observational constraints. Our resulting mock galaxy catalogues manage to reproduce the observed local galaxy luminosity function and the colour-magnitude distribution as observed by the Sloan Digital Sky Survey. They also reproduce the observed galaxy clustering properties as a function of
luminosity and colour. In order to achieve that, the algorithm also includes scatter in the halo mass - galaxy luminosity relation derived from direct SHAM and a modified NFW mass density profile to place satellite galaxies in their host dark matter haloes. Improving on general usage of the HOD that fits the clustering for given magnitude limited samples, our catalogues are constructed to fit observations at all luminosities considered and therefore for any luminosity subsample. Overall, our algorithm is an economic procedure of obtaining galaxy mock catalogues down to faint magnitudes that are necessary to understand and interpret galaxy surveys.

\end{abstract}


\section{Introduction}
\label{sec:intro}

Two of the most exciting and challenging problems in Astrophysics and Cosmology are galaxy formation (for a review see \citealt{Benson:10}) and the nature of dark energy (e.g., \citealt{Weinberg:12}). Galaxy surveys constitute one of the most powerful tools to investigate them. In order to fully exploit and interpret the observed data from galaxy surveys it is essential to produce mock galaxy catalogues since they can help in a variety of ways. They are useful to design and calibrate galaxy surveys. They can help to study selection effects, to calibrate errors and explore systematic effects, to test new techniques to measure cosmological parameters, or to calibrate cluster finders and photometric redshift estimators.

According to the current paradigm of structure formation, it is well established that galaxies are formed and reside in overdensities of the underlying dark matter field. These overdensities are called dark-matter haloes. N-body simulations and analytical models give a detailed picture of the abundance and clustering of dark-matter haloes. The link between the properties of the halo and galaxy populations is a key ingredient in order to understand galaxy formation. Galaxies form by the cooling and condensation of gas inside these dark matter haloes. Besides cooling, complicated physical properties such as star formation, merging, tidal interactions and several feedback processes, determine galaxy formation and its evolution. There are several methods to \textit{simulate} these processes and reproduce galaxy formation. These methods can be used, by themselves and together with other ingredients, to build mock galaxy catalogues. In recent years a big effort has been devoted to generate mock galaxy catalogues that reliably reproduce the observed universe, in terms of galaxy formation models, statistical approaches and computational technology.

There exist many different ways of producing mock galaxy catalogues (an incomplete list of examples includes \citealt{Cole:98}, \citealt{Bower:06}, \citealt{Croton:06}, \citealt{DeLucia:07}, \citealt{Huff:07}, \citealt{Jouvel:09}, \citealt{Cabre:11}, \citealt{Font-Ribera:12}, \citealt{Manera:12}, \citealt{Merson:13}). Galaxy mocks need to have the correct spatial clustering to be useful for cosmological purposes. The most common approach to ensure proper clustering properties when building mock galaxy catalogues is to start from an N-body dark matter simulation.  Dark matter haloes found in the simulations serve as the seed points to place galaxies. The specific recipes used to populate the dark matter haloes with galaxies differentiate the different methods. 

Three common different approaches are used to connect galaxies with dark matter haloes: hydrodynamical simulations (e.g., \citealt{Katz:92}, \citealt{Evrard:94}, \citealt{Frenk:96}, \citealt{Weinberg:97}, \citealt{Pearce:99}, \citealt{Springel:03}, \citealt{Springel:05},  and for a review, see for example \citealt{Springel:10}), semi-analytic models (SAMs) (e.g., \citealt{White:78}, \citealt{White:91}, \citealt{Kauffmann:93},  \citealt{Kauffmann:99}, \citealt{Benson:00}, \citealt{Cole:00}, \citealt{Hatton:03}, \citealt{Springel:05}, \citealt{Somerville:08}, \citealt{Benson:12}) and extensions of the halo model (see the review by \citealt{Cooray:02}), such as the conditional luminosity function (CLF) (e.g. \citealt{Peacock:00}, \citealt{Yang:03}, \citealt{vandenbosch:03}, \citealt{Cooray:05}, \citealt{CoorayMilosavljevic:05}, \citealt{Cooray:06}, \citealt{vandenbosch:07}, \citealt{Wang:10}), the halo occupation distribution (HOD) (e.g. \citealt{Jing:98}, \citealt{Benson:00}, \citealt{Seljak:00}, \citealt{Scoccimarro:01}, \citealt{Berlind:02}, \citealt{Bullock:02}, \citealt{Zheng:02}, \citealt{Berlind:03}, \citealt{Magliocchetti:03}, \citealt{Kravtsov:04}, \citealt{Zehavi:05}, \citealt{Tinker:05}, \citealt{Tinker:07}, \citealt{Zheng:09}, \citealt{DeGraf:11}, \citealt{Zehavi:11}, \citealt{Leauthaud:11}, \citealt{Krause:13}, \citealt{Rodriguez-Puebla:13}, \citealt{Pujol:2013}) or the subhalo abundance matching (SHAM) technique (e.g. \citealt{Klypin:99}, \citealt{Kravtsov:04}, \citealt{Tasitsiomi:04}, \citealt{Vale:04}, \citealt{Conroy:06}, \citealt{Vale:06}, \citealt{Guo:10}, \citealt{TrujilloGomez:11}, \citealt{Hearin:13}).

N-body simulations including hydro-dynamics have normally focused on smaller scales than those relevant for cosmological applications. Nevertheless, there have been several attempts to run large volumes including the relevant physics of the baryonic component of the universe \citep[e.g.,][]{Gottloeber:07}.  These approaches are computationally very expensive to properly develop cosmological simulations with enough resolution and volume and have not been extensively pursued.

Semi-analytic models (SAMs) attempt to simulate galaxy formation and evolution using simple analytic prescriptions to approximate star formation and feedback processes assuming that structure grows hierarchically. One can obtain spectral energy distribution of galaxies as well as many other properties of the galaxy population by modelling the formation of galaxies inside the dark matter haloes and using the mass accretion history of dark-matter haloes (merger trees). The newest SAMs models use as input dark matter halo merging history trees from large very high-resolution N-body simulations, which include the spatial and velocity distribution of the dark-matter haloes and subhaloes and therefore they naturally incorporate the clustering properties.

The basic halo model is a formalism to describe the non-linear gravitational clustering. One could say that there exist three extensions of the halo model that relate dark-matter haloes with galaxies: (i) the halo/subhalo abundance matching (HAM/SHAM) technique assumes a monotonic relation between some determined galaxy property and the mass (or the dynamics) of the halo, (ii) the halo occupation distribution (HOD) describes the galaxy bias in terms of a probability distribution that a dark-matter halo of mass $M_{h}$ contains $N_{g}$ galaxies of a given type, (iii) the conditional luminosity function (CLF) extends the HOD approach and gives the number of galaxies that resides in a halo of mass $M_{h}$ as a function of luminosity.

Building mock galaxy catalogues that precisely follow a specific observed property can be routinely achieved with the methods described above. However, in order to  help design, optimise, analyse and interpret current and future large cosmological surveys, mock catalogues need to properly reproduce the most important observed global properties of the galaxy population and how clustering depends on them. In this paper, we present an economic method to achieve this goal starting from a halo catalogue and combining HOD and SHAM techniques to generate galaxies. In particular we optimize the method to reproduce the observed luminosity function, the observed colour distributions and the global spatial galaxy clustering and its dependence on luminosity and colour.

%%%%%%%%%%  NEW

This article is the first in a series of papers describing a method to generate mock galaxy catalogues from halo catalogues. The methodology is general and can in principle be applied to any halo catalogue.  In addition to describing the method, in this paper we also focus on its calibration using properties of the local universe from Sloan Digital Sky Survey (SDSS; ~\citealt{York:00}) data. We present the application of the method to the halo catalogues of the MICE simulations. In particular, we describe the recipes used to generate the first release of the MICE galaxy catalogues, MICECAT v1.0, built from the lightcone halo catalogue. The catalogue is publicly available through the CosmoHUB web portal\footnote{http://cosmohub.pic.es}. The clustering and lensing properties of this catalogue and its associated halo catalogue have been already explored in \citet{Fosalba:13a,Crocce:13,Fosalba:13b,Hoffmann:14}, and serve as a validation of the method and the catalogue. The exact values of the implementation of the method depend on the starting halo catalogue and as such are cosmology dependent. All the values given in the paper correspond to the optimization done to fit the MICE cosmology.

%We defer an explanation of how we apply it to a lightcone including redshift evolution to the next paper in this series. 

The article is organized as follows: section \ref{sec:input} briefly presents the observational constraints that we use to calibrate the mock galaxy catalogue and the input MICE halo catalogue, section \ref{sec:method} describes in detail the method to build the galaxy catalogue and finally section \ref{sec:conclusions} presents the discussion and conclusions.

%%%%%%%%%%%%%%

%This article is the first in a series of papers describing the process of generating mock galaxy catalogues and the catalogues themselves using as input the halo catalogues extracted from the MICE simulations. The clustering and lensing properties of a catalogue realization produced with this method have been already explored in \citet{Fosalba:13a,Crocce:13,Fosalba:13b,Hoffmann:14}\footnote{The four referred articles analise the first public release of the MICE-Grand Challenge (MICE-GC) light-cone halo and galaxy catalogue, MICECAT v1.0, which used as input the light-cone of the MICE-GC N-body run simulation. The catalogue is available through the CosmoHUB portal (http://cosmohub.pic.es) and was generated following a version of the algorithm presented in this work but including redshift evolution (Castander et al. 2014 in prep.).}. This paper will focus on the calibration of the method using properties of the local universe from Sloan Digital Sky Survey (SDSS; ~\citealt{York:00}) data. The article is organized as follows: section \ref{sec:input} briefly presents the observational constraints that we use to calibrate the mock galaxy catalogue and the input MICE halo catalogue, section \ref{sec:method} describes in detail the method to build the galaxy catalogue and finally section \ref{sec:conclusions} presents the discussion and conclusions.

\section{Input}
\label{sec:input}

Our catalogue building philosophy consists on generating galaxies starting from a halo catalogue to which we apply algorithmic recipes to generate galaxies that follow a set of observational constraints. In this section, we start by describing the observational constraints we impose to generate the galaxy catalogue. The method is general in the sense that it could be applied to any initial halo catalogue. However, the actual values used in the recipes depend on the particular halo catalogue choice used. In our case, we use the halo catalogues extracted from the MICE simulations, which we describe below.

Our catalogues are intended for cosmological purposes. It is therefore imperative that the galaxies present the observed clustering properties and their dependence on observed galaxy properties. 
Unfortunately there only exist stringent observational constraints of the galaxy clustering as a function of their properties at low redshift and thus we choose to optimise our method with local observations. We use luminosity and colour as the main galaxy properties to reproduce,  although other properties could be used if wanted. The galaxy luminosity function, i.e. the mean number density of galaxies as a function of the luminosity, $\frac{dn}{dL}$, is one of the most fundamental properties of the galaxy population. In our particular implementation, we use the SDSS luminosity function computed by~\cite{Blanton:03} at z=0.1 in the $^{0.1}$r-band\footnote{The SDSS r-band filter redshifted to z=0.1.}, with best-fitting Schechter parameters $\phi_{*}=(1.49\pm 0.04)\cdot 10^{-2}h^{3}$Mpc$^{-3}$, $M_{\ast}-5\log_{10}h = -20.44 \pm 0.01$, and $ \alpha =-1.05 \pm 0.01$. Recently, we have extended our catalogue to fainter luminosities and have used the luminosity functions extended to fainter values presented in~\cite{Blanton:05b} for calibration\footnote{Given that it is not feasible in practice for simulations sampling large cosmological volumes to have good mass resolution, and given the method we use to assign galaxy luminosities, the faintest galaxies in the MICECAT v1.0 catalogue have an absolute magnitude in the $^{0.1}$r-band of $M_{r}\lesssim-18.8$ and therefore the~\cite{Blanton:03} LF covers the range of luminosities in our catalogue. In Castander et al. 2014 in prep., we force our algorithm and we decrease by five magnitudes the minimum galaxy luminosity making the catalogue follow the SSDS luminosity function of extremely low luminosity galaxies presented in \citealt{Blanton:05b}.}. We also use SDSS galaxy clustering information. In particular we calibrate to the results of the projected 2-point correlation function of galaxies as a function of luminosity thresholds, $w_{p}(r)(>L_{r})$, presented by~\cite{Zehavi:11} (their Table~8); as well as to the projected 2-point correlation function of galaxies as a function of colour (their Tables~9 and 10). Finally, we make the mock galaxy catalogues follow the global SDSS $(^{0.1} g-^{0.1} r)$ colour-distribution as a function of the absolute magnitude in the $^{0.1}$r-band~\citep{Blanton:05c}. The distribution of rest-frame $(g-r)$ (and $(u-r)$) colours at a fixed absolute magnitude in the r-band is normally adequately modelled as the sum of two Gaussian components (e.g., \citealt{Baldry:04}, \citealt{Blanton:05}).

The particular implementation of the mock galaxy catalogues presented here is optimized using the Marenostrum Institut de Ci\`encies de l'Espai simulations\footnote{http://maia.ice.cat/mice} (MICE; \citealt{Fosalba:08}). The MICE simulations are a suite of dark-matter N-body simulations carried out at the Marenostrum supercomputer at the Barcelona Supercomputing Center\footnote{http://www.bsc.es}. The main goal of the MICE simulations is to study the formation and evolution of structure at very large scales and also to determine how well future astronomical surveys can answer the current cosmological questions. All MICE runs assume a flat concordance $\Lambda$CDM model with parameters $\Omega_{m}=0.25$, $\Omega_{\Lambda}=0.75$, $\Omega_{b}=0.044$, $h=0.7$ and the spectral index $n_{s}=0.95$. The linear power spectrum is normalized to yield $\sigma_{8}=0.8$ at $z=0$. While the catalogues generated from the MICE simulations assume a particular cosmology, the method presented in the paper is general and can be optimized for other simulations with different cosmological parameters. The suite of MICE simulations sample a wide range of cosmological volumes, from tenths to hundreds of (Gpc/h)$^{3}$, equivalent to the range of volumes surveyed by the SDSS main sample, the Dark Energy Survey (DES\footnote{http://www.darkenergysurvey.org}) and up to the one to be sampled by Euclid\footnote{http://sci.esa.int/euclid/}~\citep{Laureijs:11}. 
The mass resolutions used range from $3\cdot10^{12}h^{-1}$M$_{\bigodot}$ down to $3\cdot10^{8}h^{-1}$M$_{\bigodot}$.

We use the newest MICE run, the  Grand Challenge (GC) simulation \citep{Fosalba:13a,Crocce:13,Fosalba:13b}, to calibrate the algorithm and to derive statistical errors. In particular we use the snapshot at $z=0$. The MICE-GC snapshots cover a large volume with a box-size of $L_{box}=3072$ Mpc/h. We divide the whole volume into $10^{3}$ sub-boxes of $L_{box}=307.2$ Mpc/h to have more manageable sizes to optimize the method and compute statistics. The MICE halo catalogues are derived using the public halo finder code available at the N-body Shop\footnote{http://www-hpcc.astro.washington.edu/}. From the N-body dark matter particles we compute haloes running the friend-of-friends (FoF) code with a linking length b=0.2 times the mean inter particle distance. We retain haloes formed by linked particle associations with as few as 10 particles ($N_{p}^{min}=10$). As explained below, our luminosity assignation is based on abundance matching between halo mass and galaxy luminosity. Therefore, pushing to lower halo masses allow us to generate fainter galaxies. For modern deep cosmology surveys that reach faint magnitudes, it is necessary to push to faint luminosities in order to generate catalogues that are complete. It is unclear whether FoF groups of such few particles represent virialized haloes. However, our objective is to generate galaxies that follow observed clustering patterns and our method does so. We leave a detailed investigation about these resolution issues and their effects on clustering to further work.
At the other end of the mass range, the FoF halo finding linking length used of $b=0.2$ may cause over-linking for the most massive structures. Given their low number, their influence in the statistical properties of the whole population is negligible. Nevertheless, studies of the most massive haloes using the MICE-GC simulation should take this fact into account.

\cite{Crocce:10} studied the abundance of massive haloes in the MICE simulations\footnote{The MICE-GC run was not available at that time.}. They provided a re-calibration of the halo Mass Function over 5 orders of magnitude in mass ($10^{10} < $M$  (h^{-1} M_{\odot})<10^{15}$), that accurately describes its redshift evolution up to $z=1$. The halo mass is computed taking into account the Warren correction (\citealt{Warren:06}), given by the following equation: 

\begin{equation}\label{eq:halo_mass_cap4}
M_h^{W}=m_{p}(N_{p}(1-N_{p}^{-0.6}))
\end{equation}
that corrects a systematic problem that FOF halo masses suffer when haloes are sampled by relatively small number of particles. This correction was tested for the MICE cosmology in~\cite{Crocce:10}, where it was proved suitable. Anyhow, this correction is inconsequential in our method of assigning galaxy luminosities based on abundance matching. We have generated catalogues with and without it, without any difference. The MICECAT v1.0 presented in this paper includes it, though. \cite{Crocce:13} present a detailed statistical description of the MICE-GC halo population, which we do not repeat here. For each FoF halo we compute its mass, position and velocity of its centre of mass. We use these halo properties as the starting point to generate the mock galaxy catalogues.

\section{Method}
\label{sec:method}

This section describes the method we follow to build mock galaxy catalogues. The algorithm basically combines the HOD model and the SHAM technique to assign galaxy luminosities. We also describe the way we place galaxies inside their host dark matter haloes which approximately follows an NFW profile~\citep{Navarro:97}. We assign colours to the galaxies in a similar fashion as in~\cite{Skibba:09}, which we later use to assign them spectral energy distributions (SEDs). We will describe that process in forthcoming papers. We calibrate the algorithm using the halo catalogue extracted from the snapshot at z=0 of the MICE-GC run and the observational constraints presented in section~\ref{sec:input}. As we will describe below, our catalogues fit the galaxy luminosity function and the colour magnitude diagram by construction. In order to fit the galaxy clustering, we generate catalogues modifying the parameters of the method and compare their clustering to the observed clustering measurements of~\cite{Zehavi:11}, hereafter Z11. We minimize the difference to obtain the parameters of the method that best match the observed clustering.

The overall concept of the method is simple. However, we cannot manage to reproduce the galaxy properties in detail for the whole catalogue with these simple prescriptions and need to perform slight modifications and alterations to the standard recipes in order for them to work properly. We attribute this to the complex diversity of the galaxy population that it is hard to capture with simple recipes.

This section is divided in four subsections. The first one explains the way we assign luminosities to the galaxies in order to follow the observed luminosity function. The second describes how we place galaxies inside the haloes to reproduce the observed luminosity dependence of galaxy clustering. The third subsection explains how we assign galaxy velocities and the last one shows how we assign galaxy colours in order to match the observed (g-r) .vs. $M_{r}$ diagram and reproduce the clustering properties as a function of colour. Although we describe the method following these four broad subsections, we note that they are intertwined in our implementation.

\subsection{Galaxy luminosities}
\label{sec:galaxy_luminosity}

The luminosity function is one of the most important properties of the galaxy population. Our method uses the subhalo abundance matching (SHAM) technique to assign luminosities to the galaxies. Therefore, the resulting galaxy population follows the empirical observational constraint by construction. The common approach when using the SHAM technique is to extract the substructure inside haloes and relate the subhalo mass (or their velocity dynamics) with some galaxy property, such as the stellar mass or the luminosity. This procedure implies that one needs to properly sample the haloes to resolve their internal substructure. This is however not feasible in practice for simulations sampling large cosmological volumes where the mass resolution does not allow to properly sample subhaloes. We follow a more economical approach that allow us to reach much fainter into the luminosity function for a simulation of a given mass resolution. Instead of resolving the internal structure of the halo with subhaloes that are matched to galaxies, we compute the expected number of satellites in each halo depending on its mass given by the Halo Occupation Distribution (HOD) and then this number is used in the abundance matching technique. The method becomes hybrid as one needs to fit both, the HOD and the SHAM, simultaneously. In fact, one needs to simultaneously fit the internal structure of the haloes as well, as we will see later.  This section describes this hybrid method to assign galaxy luminosities using the HOD model and the SHAM technique. It guarantees that the resulting luminosity galaxy distribution of the mock catalogue follows the observed luminosity function.

\subsubsection{Halo Occupation Distribution modelling}
\label{sec:HOD_modelling}

The HOD model relates dark matter and galaxies. This relation, which as a whole can be referred to as galaxy bias, is fully defined by (i) the probability distribution that a halo with a given mass contains a number of galaxies of a given type, i.e. $P(N_{g} | M_{h})$, (ii) a relation between the galaxy and the dark matter spatial distributions and (iii) a relation between the galaxy and dark matter velocity distribution. This subsection presents the functional form used to describe $P(N_{g} | M_{h})$. \cite{Kravtsov:04} studied the HOD and 2-point correlation function of galaxy-size dark-matter haloes using high-resolution dissipationless simulations and they found that the probability for a halo of mass, $M_{h}$, to host a number of subhaloes, $N_{sh}$, is similar to that found in semi-analytic and N-body + gas dynamics studies. They showed that the HOD can be thought as a combination of the probability for a halo of mass, $M_{h}$, to host a central galaxy, $N_{cen}$, and the probability to host a given number of satellite galaxies, $N_{sat}$. 

The HOD model is often used to understand the clustering properties of a galaxy sample in terms of how galaxies populate haloes. Different authors have computed the best-fitting HOD parameters to match different observations. The general idea is to choose an HOD parametrization and find the set of HOD parameters that best fit at the same time the mean number density of galaxies as a function of luminosity and the clustering (e.g., the 2-point projected correlation function) as a function of luminosity for different galaxy samples (e.g., \citealt{Zehavi:05}, \citealt{Zheng:09}, \citealt{Zehavi:11}, \citealt{Krause:13}, \citealt{Richardson:13}). Normally for luminosity threshold samples five parameters are needed to properly describe the observations. \cite{Zheng:05} provide a parametrization that is widely used which includes two parameters to describe the occupation function of central galaxies with an $\erf$ function (their equation~1) and three parameters for satellite galaxies with a power law (their equation~3).

Our purpose here is somewhat different than the general HOD usage. We are interested in assigning luminosities to the galaxies belonging to haloes of known mass while normally the HOD prescription is used for the opposite purpose to understand the mass of the haloes that galaxies of an observed luminosity populate. For normal luminosity threshold samples that do not reach faint luminosities, one cannot accurately describe the clustering if one imposes a sharp cut in halo mass corresponding to the luminosity threshold. The galaxies with luminosity values close to the threshold, which are mostly centrals, need to populate haloes of a range of halo masses to reproduce observations. This behaviour is analogous to what happens in the SHAM technique, where, as we will see later, one needs to include scatter between galaxy luminosities and halo masses to match observations. 

In our case we start from a halo mass limited sample to which we want to assign luminosities to its galaxy constituents.
We have chosen a simple HOD functional form in which each halo contains one central galaxy and the average number of satellites galaxies in a halo is a power-law of the halo mass.
Our simple HOD functional form thus only contains three free parameters, $M_{min}$, the minimum halo mass for a halo to host a central galaxy; $M_{1}$, the halo mass at which a halo contains on average one satellite galaxy; and $\alpha$, which is the high-mass slope of the satellite galaxy mean occupation. We assume that haloes more massive than $M_{min}$ host one central galaxy,

\begin{equation}\label{eq:Ncen_exp}
<N_{cen}> = 1	\ \ \ \ \mbox{if} \ \ M_h \geq M_{min}
\end{equation}
 and the number of satellite galaxies follows a Poisson distribution with the mean following the expression:

\begin{equation}\label{eq:Nsat_exp}
<N_{sat}> = \left[\frac{M_h}{M_{1}} \right]^{\alpha}	\ \ \ \ \mbox{if} \ \ M_h \geq M_{min}.
\end{equation}
We assume that haloes less massive that $M_{min}$ do not contain any galaxy, neither central nor satellite.

Note that here we deviate from standard HOD clustering analysis of observed luminosity threshold samples which require a 
dispersion of the halo masses at the luminosity threshold (e.g, the $\erf$ function in the \cite{Zheng:05} parametrization), while we use a sharp luminosity threshold for a halo mass threshold given by the one-to-one correspondance of central galaxies and haloes (e.g., each halo contains one central galaxy). The reason why our approach is appropriate to reproduce clustering properties follows from the way the halo bias changes as a function of halo mass, which is shown in Figure~\ref{fig:bias_Halo_Mass_r_16.9}. For high halo mass values the change in the bias is steep with halo mass while the bias is almost constant for low halo masses. In normally used observed luminosity threshold samples, the luminosity threshold is ``relatively'' bright corresponding to halo masses where the bias is still a steep function of its mass. Normally for a ``one central galaxy for each halo'' approach, which yields a sharp halo mass threshold, the clustering is stronger than observed. One is therefore forced to reduce the mass of the halo
of the lowest luminosity central galaxies to reduce the clustering strength\footnote{The argument here is somewhat oversimplified as there is another degree of freedom to play with given by the occupation number of the satellite galaxies. Nevertheless it is still qualitatively valid.}.  The $\erf$ function of the~\cite{Zheng:05} parametrization serves this effect. This is the same argument why in SHAM techniques one needs to implement scatter in the halo mass-luminosity relation to reduce the clustering strength and match observations. In our case, we build catalogues down to halo masses ($\sim10^{11} M_{\odot}/h$) where their bias is almost constant and does not depend on the halo mass. Therefore our clustering strength does not depend whether we assign a fuzzy or a fixed luminosity threshold to our fixed halo mass threshold. We choose a fixed luminosity threshold for convenience so that our resulting catalogue will be luminosity limited.

\begin{figure}
	\centering
	\includegraphics[width=8cm]{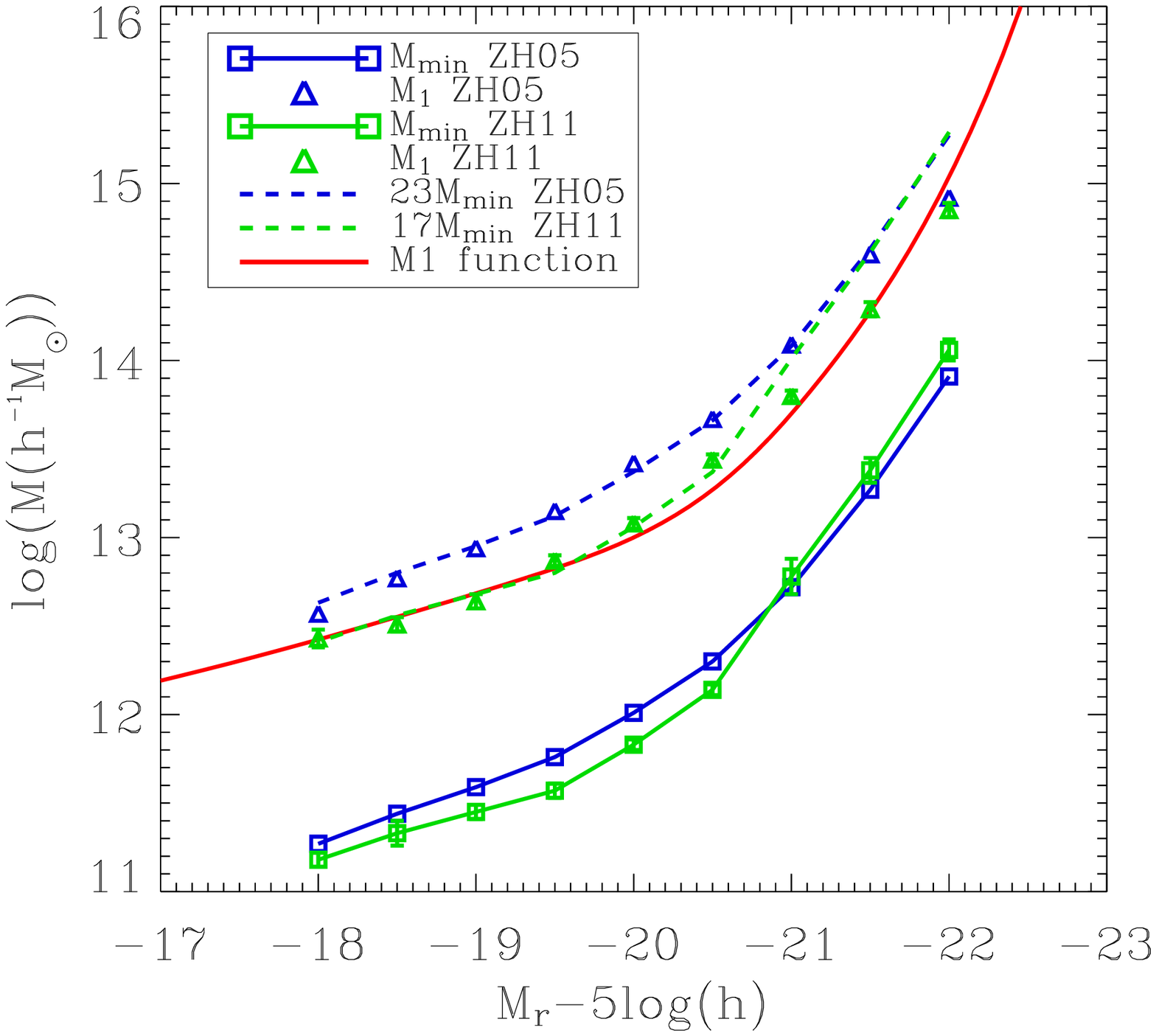}
	\caption{\small{$M_{min}$ and $M_{1}$ HOD parameters as a function of the luminosity threshold of galaxy samples. The red solid line is the fit we set to model $M_{1}$ as a function of $M_{min}$. The blue squares and triangles are the HOD parameters found by \citet{Zehavi:05} while the green squares and triangles refer to the parameters found by \citet{Zehavi:11}. The dashed blue line is the $M_{min}-M_{r}$ found by \citet{Zehavi:05} scaled by a factor of 23, and the dashed green one is the same relation found by \citet{Zehavi:11} but scaled by a factor of 17.}}
	\label{fig:function_M1}
\end{figure}

Returning to our HOD parametrization, we adopt a somewhat different philosophy than normally found in the literature, allowing one parameter to be a function of mass instead of just one value. While for different luminosity threshold samples one can choose a set of HOD parameters to fit observations, we find that we cannot find a set of parameters that can fit the observations at all the luminosities that our catalogue spans. In particular our implementation assumes the parameter $\alpha=1$, consistent with values found in the literature \citep[e.g.,][]{Kravtsov:04, Zehavi:11}.
Instead of using discrete values for the parameter $M_{1}$, we model the dependence of $M_{1}$ on $M_{min}$. The factor $f_{M_{1}}$, which multiplies $M_{min}$ to obtain $M_{1}$, varies with halo mass (or luminosity). We have arbitrarily chosen to express the $f_{M_{1}}(M_{halo})$ factor function as a combination of $\tanh$ functions that smoothly transition from constant values when varying the halo masses. This function resembles the changing values of the parameter when matching different observed samples at given luminosity thresholds. In principle any functional form could be used as long as it is continuous and varies smoothly with low values of its derivative with respect to halo mass. Otherwise it introduces discontinuities in the luminosity function in our hybrid method to generate galaxy catalogues. 
In the end we find that the expression in equation~\ref{eq:facM1_function} produces catalogues that are a reasonable fit to observations:

\begin{multline}\label{eq:facM1_function}
f_{M_{1}}=0.5((a_{1}-a_{2})\tanh{(s_{1}(b_{1}-\log{M_{h}}))}+ \\ +(a_{3}-a_{2})\tanh{(s_{2}(\log{M_{h}}-b_{2}))}+(a_{1}+a_{3}))
\end{multline}
where $a_{1}=25.0$, $a_{2}=11.0$, $a_{3}=14.0$, $b_{1}=11.5$, $b_{2}=12.5$, $s_{1}=2.0$ and $s_{2}=2.50$. The $a_i$ parameters give the value of the $f_{M_{1}}$ factor. The $b_i$ parameters give the halo mass values where the transitions between different $a_i$ values occur and the $s_i$ parameters give the abruptness of the transitions. Note that this function is not optimized in isolation but in conjunction with many others to produce a catalogue that matches observations. As we will see in more detail later, we use this function to generate the number of satellites per halo as a function of their halo mass. Then we generate luminosities for these galaxies with abundance matching and finally we assign positions to these galaxies assuming an NFW profile. Finally, we compute the clustering of the catalogue as a function of luminosity and compare it to observations. We change the values of the parameters in this function together with the abundance matching scatter and the NFW parameters to minimize the $\chi^2$ of the comparison of the galaxy catalogue clustering to observations. Figure~\ref{fig:function_M1} shows the resulting $M_{1}$ HOD parameter as a function of the luminosity (solid red line) compared to the best-fitting HOD parameters $M_{min}$ (squares) and $M_{1}$ (triangles) found by~\cite{Zehavi:05} (in blue) and~\cite{Zehavi:11} (in green) for different luminosity threshold galaxy samples. \cite{Zehavi:05} found that their $M_{1}$ values could be approximated by their values of $M_{min}$ multiplied by a constant factor of 23. This approximate $M_{1}$ fit is represented in Figure~\ref{fig:function_M1} by a dashed blue line. In the same way, Z11 found that a factor 17 could well reproduce their values of $M_{1}$ given their values of $M_{min}$. This fit by Z11 is presented by a dashed green line in  Figure~\ref{fig:function_M1}. We choose to show these relations as they follow the same parameterization we use for $M_{1}$ as a function of $M_{min}$ and show that a single value for this factor is not a good approximation to the actual values of $M_{1}$ as a function of halo mass at least in the case of Z11.
The large difference in the data points between~\cite{Zehavi:05} and Z11 comes partly from the different SDSS data and mainly because of the different assumed cosmological parameters in both studies. In particular because~\cite{Zehavi:05} assumed $\sigma_{8}=0.9$ and Z11 assumed $\sigma_{8}=0.8$. Our fit to model $M_{1}(M_{r})$ is quite similar to the values found by Z11. This is also expected since the MICE cosmological model and the one assumed in their HOD modelling are similar.

\subsubsection{Subhalo Abundance Matching modelling}
\label{sec:SHAM_modelling}

The other ingredient we use in our method is the Subhalo Abundance Matching technique which we combine with the HOD prescription. First, we proceed to assign luminosities to the central galaxies. Given an HOD model and the halo mass function of the halo catalogue, one can compute the cumulative mean number density of galaxies that inhabit haloes more massive than a certain value splitting the galaxies into centrals and satellites, 

\begin{equation}\label{eq:gal_cen_sat}
n_{gal}\left(>M_{min}\right)=n_{cen}\left(>M_{min}\right)+n_{sat}\left(>M_{min}\right),
\end{equation}
using the following expressions:

\begin{equation}\label{eq:cmfcen}
n_{cen}\left(>M_{min}\right)=\int_{M_{min}}^{\infty}N_{cen}(M_{h})\frac{dn}{dM_{h}}dM_{h}
\end{equation}

\begin{equation}\label{eq:cmfsat}
n_{sat}\left(>M_{min}\right)=\int_{M_{min}}^{\infty}N_{sat}(M_{h})\frac{dn}{dM_{h}}dM_{h}
\end{equation}

Complementary, we compute the mean number density of galaxies with luminosity brighter than a certain value, $n_{gal}(<M_{r})$, using the best-fitting Schechter function derived from observations by~\cite{Blanton:03} which we complement with the extension of~\cite{Blanton:05b} when extending our fits to low luminosities:

\begin{equation}\label{eq:clfgal}
n_{gal}(>L_{r})=\int_{L_{r}}^{\infty}\frac{dn}{dL_{r}}dL_{r}
\end{equation}

Following the SubHalo Abundance Matching (SHAM) technique,
we are able to derive a relation between mass and galaxy luminosity, $M_{halo}-L_{gal}$, equalling both cumulative functions: 

\begin{equation}\label{eq:SHAM}
n_{gal}(>M_{min})=n_{gal}(>L_{r})
\end{equation}

As stated before, in our case we impose to have one central galaxy per halo, $N_{cen}(M_{h}) = 1$ if $M_{h} \ge M_{min}$ and the number of satellites per halo, $N_{sat}(M_{h}$), follows a Poisson distribution with mean given by equation~\ref{eq:Nsat_exp}. When building a mock catalogue with large number of galaxies, it is slow and inefficient to evaluate equation~\ref{eq:SHAM} for each galaxy. Instead we precompute multiple values of halo mass and luminosity pairs that we fit with a spline that we later use as our $M_{halo}-L_{gal}$ relation\footnote{We found that we needed functions of at least five parameters to fit this relation in the range of halo masses and luminosities of our catalogues. In the end we decided to use a more flexible spline fitting.}. Note that this relation is cosmology dependent as the mass function varies with cosmology.

We assume that central galaxies follow the $M_{halo}-L_{gal}$ relation, and thus derive the central galaxies luminosities directly from the halo masses they populate. This procedure is not strictly correct as we use a relation found using abundance matching techniques counting all galaxies and not just centrals. However, we find that this procedure produces catalogues that can match observations and therefore avoid the complications and longer running times of more adequate procedures. 

We then assign luminosities to satellite galaxies drawing them from their cumulative luminosity function. 
We can compute the contribution of central galaxies to the total galaxy luminosity function by introducing the  $M_{halo}-L_{gal}$ relation into equations \ref{eq:Ncen_exp} and \ref{eq:cmfcen},

\begin{equation}
n_{cen}(>M_{min}(L_{gal}))=\int_{M_{min}(>L_{gal})}^{\infty}\frac{dn}{dM_{h}}dM_{h}
\end{equation}
and subtract it from the observed total cumulative galaxy luminosity function to obtain the satellite cumulative luminosity function:
\begin{equation}\label{eq:nsatL}
n_{sat}(>L_{r})=n_{gal}(>L_{r})-n_{cen}(>L_{r}),
\end{equation}

This function is the global satellite luminosity function. We further assume that one can draw randomly from this distribution to assign luminosities to satellite galaxies with the additional restriction that they cannot be much brighter than the central galaxy of the halo. This assumption produces individual halo luminosity functions that are in reasonable agreement with galaxy cluster luminosity functions.

In more detail, we compute the number of satellite galaxies inside each halo given its mass drawing a realization of the HOD mean occupation number  (equation~\ref{eq:Nsat_exp}). We end up with the total number of satellite galaxies in the catalogue, $N_{sat}$, summing the satellites in all the individual haloes. We generate $N_{sat}$ luminosities by randomly sampling the cumulative satellite luminosity function, $n_{sat}(>L_{r})$ (equation~\ref{eq:nsatL}). Assigning the generated $N_{sat}$ luminosities to the satellite galaxies, the global distribution of luminosities will follow the observed luminosity function. Given that observationally clusters are found to generally have a central dominant galaxy that is the most luminous in the cluster\footnote{We expect that, in general, the most luminous galaxy in a halo is its central galaxy, although there are several works that claim that a non negligible number of haloes does not contain the brightest galaxy at its centre (e.g., \citealt{Skibba:11}).}, we include another assumption when assigning satellite luminosities. We enforce satellite galaxies not to be brighter than a certain fraction of the luminosity of the central galaxy. In particular we assign one of the sampled luminosities if $L_{sat}\leqslant 1.05 \cdot L_{cen}$. We have tried values between 1.05 and 1.30 for the factor between the maximum allowed satellite luminosity and the luminosity of the central galaxy in the same halo. For each value of that factor we have generated a catalogue realization and computed a diagonal-$\chi^2$ between the clustering as a function of luminosity of that catalogue with respect to observations (Z11). We compute 
\begin{equation}\label{eq:chi2}
\chi^2 = (w_p^{cat} - w_p^{obs})^T C^{-1} (w_p^{cat} - w_p^{obs})
\end{equation}
where $w_p^{cat}$ ans $w_p^{obs}$ are the vectors of the projected correlation functions for the catalogue and observations respectively computed at scales $r_p$ and $C$ is the error covariance matrix. As the tables in Z11
do not provide the full covariance matrix values, we only take the diagonal elements, and therefore refer to $\chi^2$ as diagonal-$\chi^2$. The goodness of the fits are similar for all the values of the factor tried, with differences of the diagonal-$\chi^2$ per degree of freedom lower than 0.5.
We have finally somewhat arbitrarily opted for 1.05. Once a sampled luminosity is assigned to a satellite galaxy, it cannot be assigned again to another satellite galaxy. Following this method one gets to a point in which no sampled satellite luminosity fulfills the constraint. It is very important to remark that the number of objects left is very small, less than one in ten thousand, although it depends on the luminosity range sampled. In these rare cases we assign the minimum luminosity of the catalogue to the satellite\footnote{We have explored other possibilities but given the small number of objects, it does not make any difference.}.

%The method described in this section guarantees that the constructed mock galaxy catalogue follows the observed luminosity function.

So far, we have described the method that guarantees that the constructed mock galaxy catalogue follows the observed luminosity function. We now turn our attention to how the galaxy clustering behaves as a function of luminosity.

\subsubsection{Scatter in the halo mass central luminosity relation}
\label{sec:scatter}

As we will discuss below, the main contribution to the galaxy clustering in the linear regime comes from the spatial distribution of the dark-matter halo population, which is the same as the central galaxies in our catalogue. If we follow the SHAM method described to assign luminosities and  place central galaxies at the centre of the haloes, the clustering of the mock more luminous galaxies presents a larger amplitude than observations no matter the HOD model we assume for the MICE-GC simulation and its associated cosmology. 
The cause for this clustering excess can be understood in the way the SHAM method assigns luminosities to galaxies. The most luminous galaxies end up in the most massive haloes with the SHAM technique and the clustering amplitude of those most massive haloes is stronger than the clustering of the observed galaxy sample whose luminosity threshold corresponds to the SHAM matched halo mass threshold. The way to reduce the clustering of the luminosity threshold sample is to have those galaxies in less massive clusters. In our hybrid implementation of the HOD and SHAM method, the luminosity and halo assignment of the central galaxies is set by the SHAM technique and naively one would think that the way to reduce the clustering would be to include satellites with a lower clustering amplitude to reduce the overall clustering strength.  This is not possible because of the way our hybrid technique works. If we introduce satellite galaxies, the total cumulative number of galaxies as a function of halo mass, $n_{gal}(>M_{min})$ at a certain halo mass $M_{h}$, will be larger than only having central galaxies. In the SHAM technique one has to maintain constant the mean number density of galaxies given by the luminosity function, therefore the threshold in the total cumulative number of galaxies as a function of halo mass increases to more massive haloes, increasing the overall clustering amplitude. Therefore it is not possible to decrease the clustering of the total galaxy catalogue by including satellite galaxies in our hybrid HOD and SHAM method. The way to populate less massive haloes is to introduce scatter in the halo mass - galaxy luminosity relation. Scatter in the observable-halo mass relation has also been included previously in other studies \cite[e.g.,][]{Tasitsiomi:04,More:09b,Yang:09,Behroozi:10,Moster:10,TrujilloGomez:11,Leauthaud:12}. 
The effect of introducing scatter in the luminosity-halo mass relation is to reduce the clustering amplitude of luminosity threshold samples. This effect is most important for the brightest, and more biased, galaxies \cite[e.g.,][]{Reddick:13}. Although we argue the need for this scatter in clustering terms, the underlying reason of this scatter is due to the complexity of the galaxy formation processes.

\begin{figure}
	\centering
	\includegraphics[angle=0, width=8cm]{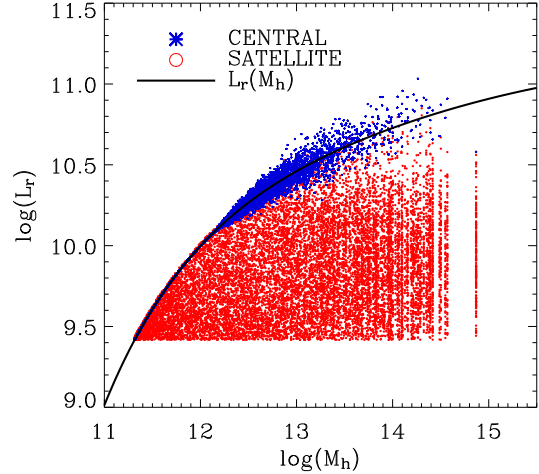}
	\caption{\small{Relation between central galaxy luminosities and their host halo mass showing the scatter applied in the halo mass-central luminosity relation. In the figure a sample of central and satellite galaxies from a simulation of box-size $L_{box}=307.2$ Mpc/h is shown. The scatter is applied for galaxies brighter than $M^{*}_{r}=-20.44$}.}
	\label{Mh-Lcen_rel}
\end{figure}

As previously mentioned in section~\ref{sec:galaxy_luminosity}, we assign central galaxy luminosities employing the relation between luminosity and halo mass found with abundance matching. Now, we add a log-normal scatter to the luminosity. The effect can be visualized in Figure~\ref{Mh-Lcen_rel}. The black line shows the initial luminosities of the central galaxies with the SHAM derived relation and the blue points the resulting central luminosities once the scatter is applied (only for the most luminous galaxies as we will see later). If we draw a luminosity threshold sample with a horizontal line in the plot, it corresponds to a sharp halo mass threshold sample with the abundance matching relation. If we include scatter, central galaxies of the same luminosities now populate haloes of a larger range of masses, and in particular many belong to less massive haloes than the previous mass threshold. As there are more centrals scattering to lower halo masses than larger halo masses, the overall effect is a lower mean halo mass for the sample and therefore a lower clustering amplitude for the resulting central galaxy sample as shown in Figure~\ref{fig:bias_Halo_Mass_r_16.9}. 
For example, if we choose a luminosity threshold sample at $M_r<-22.0$ or equivalently at $\log\,L>10.7$, that would correspond to a halo mass threshold sample of $M_{min}>13.85$ (see Figure~\ref{Mh-Lcen_rel}), which has a bias of $b\sim2.2$ according to Figure~\ref{fig:bias_Halo_Mass_r_16.9}. If instead we include scatter, then our $M_r<-22.0$ sample corresponds to halo masses $M_{min}>13.10$ with a bias of  $b\sim1.4$.

We include scatter in the halo mass-luminosity relation computing an ``unscattered" luminosity function ($\Phi(\log L')_{unscat}$ in equation~\ref{eq:LF_scatter}) that after introducing the scatter results in a cumulative luminosity function of the catalogue that matches the observed cumulative luminosity function. This is basically a deconvolution problem

\begin{equation}\label{eq:LF_scatter}
\Phi(\log L)_{obs} = \int^{\infty}_{-\infty} \Phi(\log L')_{unscat} \;G(\log L' -\log L) d\log L' 
\end{equation}
where the function $G$ is a Gaussian in the logarithm of the luminosities (or a log-normal in luminosity) with scatter $\sigma_{\log L}$.  To compute $\Phi(\log L')_{unscat}$ we follow a method similar to the one described in \cite{Behroozi:10}.  We start from the observed luminosity function to which we apply log-normal scatter in the luminosity to obtain a scattered LF. We subtract these LFs to obtain a first approximation of the unscattered luminosity function to which we apply the log-normal scatter to obtain a luminosity function that we compared to the observed one. We reiterate this process until our ``scattered'' LF is a good fit to the observed one, retaining the unscattered luminosity function. We then proceed to compute the $M_{halo}-L_{gal}$ relation using the same SHAM technique but we now use the ``unscattered" luminosity function and the cumulative number density of objects, central and satellite galaxies, $n_{gal}(>M_{min})$ (equation~\ref{eq:gal_cen_sat}). Note, however, that we deviate from previous work because we only apply scatter for luminosities above a certain threshold. As explained before, the reason to introduce scatter in the halo mass-luminosity relation is to lower the clustering amplitude making luminous galaxies to reside in lower mass haloes. Figure~\ref{fig:bias_Halo_Mass_r_16.9} shows how this method is effective at large halo masses given the steep dependence of bias on halo mass. However, at low halo masses the relation becomes flat and adding scatter to the galaxy luminosities will not change their clustering amplitude as they will populate haloes with the same bias. Given that we start from a halo mass limited sample, not including scatter at low halo masses produces a galaxy catalogue with a sharp threshold in luminosity. We have thus chosen not to include scatter at low luminosities, although one expects that the same physical processes affecting bright luminosity central galaxies will also somehow affect low luminosity centrals. We note, nevertheless, that we have an intrinsic scatter in the halo masses at our low mass end due to the few number of particles composing a low mass halo. In practice, we assign central luminosities following the relation $M_{halo}-L_{gal}$ and we apply scatter to the luminosity if it is larger than a given threshold $\log{\left( L_{r}^{thres}\right)}$. In order to smoothly transition from luminosities where we apply the full scatter to where we apply no scatter we progressively change the scatter over a range $\Delta_{\log \left( L_{r}\right)}$. In our method, if $\log{\left( L_{r}^{thres}\right)} < \log{\left( L_{r}^{cen}\right)} \leqslant \log{\left( L_{r}^{thres}\right)} +\Delta_{\log \left( L_{r}\right)}$ then we calculate the scattered luminosity as:

\begin{multline}\label{eq:scatter1}
\log{\left( L_{r}^{cen}\right)} = \log{\left( L_{r}^{cen}\right)}+ \\ +u_{1}\sigma_{\log\left( L_{r}\right)}\left( \log{\left( L_{r}^{cen}\right)}-\log{\left( L_{r}^{thres}\right)}\right)/\Delta_{\log \left( L_{r}\right)}
\end{multline}
and if $\log{\left( L_{r}^{cen}\right)} > \log{\left( L_{r}^{thres}\right)}+\Delta_{\log \left( L_{r}\right)}$:
\begin{equation}\label{eq:scatter2}
\log{\left( L_{r}^{cen}\right)} = \log{\left( L_{r}^{cen}\right)}+u_{1}\sigma_{\log \left( L_{r}\right)}
\end{equation}
where $u_{1}$ is a Gaussian distributed random number with zero mean and standard deviation equal to one, and $\sigma_{\log\left( L_{r}\right)}$ is the maximum amplitude of the scatter. We have tried different luminosity ranges where to apply the scatter and different values for $\sigma_{\log\left( L_{r}\right)}$. We find acceptable fits to the clustering luminosity dependence of the catalogue (Figure~\ref{fig:wp_th_bins_plot}) if we start applying scatter at the characteristic luminosity of the LF, $M^{*}_{r}<-20.44$ or $\log{\left( L_{r}^{thres}\right)} >10.08$ which corresponds to $\log\, M_{halo}>12.11$, smoothly transition to full scatter over a range of one magnitude or $\Delta_{\log \left( L_{r}\right)}=0.4$ and use a value for the scatter of $\sigma_{\log\left( L_{r}\right)} = 0.15$, consistent with previously found values in other studies~\citep{More:09b,Yang:09,Behroozi:10,Reddick:13}. To help visualize how scatter is applied, we present in Table~\ref{table:scatter_values} the value of the scatter applied to the logarithm of the luminosity of central galaxies depending on its luminosity. We also show the absolute magnitude and halo mass corresponding to that unscattered luminosity.

\begin{center}
\begin{table*}
\scriptsize
\centering
% use packages: array
\begin{tabular}{l|cccccccc}
\hline
$M_r^{central}$ & -19.0 & -19.5 & -20.0 & -20.5 & -21.0 & -21.5 & -22.0 & -22.5  \\
\hline
$log(L_{central})$ & 9.50 &  9.70  &   9.90 & 10.10 & 10.30 & 10.50 & 10.70 & 10.90 \\
\hline
$log(M_{halo})$ & 11.41 & 11.60 & 11.84 & 12.15 & 12.55 & 13.11 & 13.91 & 14.94  \\
\hline
$log L\; scatter$ & 0.0 & 0.0 & 0.0 & 0.02 & 0.08 & 0.15 & 0.15 & 0.15  \\
 
\hline
\end{tabular}
\caption{\small{Values of the scatter applied to the luminosities of central galaxies, together with the correspondance between absolute magnitude, luminosity and halo mass.}}
\label{table:scatter_values}
\end{table*}
\end{center}

\begin{figure}
	\centering
	\includegraphics[angle=0, width=8cm]{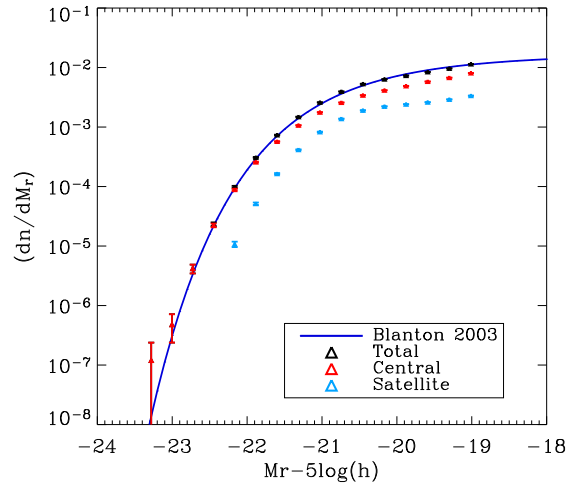}
	\caption{\small{Luminosity function of a $L=302.7$ Mpc/h sub-box of the mock galaxy catalogue. The blue solid line is the best-fitting Schechter function to the SDSS data shifted to z=0.1. The black, blue and red triangles are the luminosity function of total (central+satellite), central and satellite galaxies of the catalogue respectively. The error bars are derived as Poisson errors.}}
	\label{fig:LF_good}
\end{figure}

One of the goals of our mock catalogue is that it reproduces the observed luminosity function. The abundance matching technique automatically ensures that goal. However, we need to check whether we still reproduce the LF after applying scatter to the luminosities. For that purpose, Figure~\ref{fig:LF_good} shows the luminosity function of the generated mock galaxy catalogue using a sub-box with size $L_{box} = 307.2$ Mpc/h of the snapshot at $z=0$ of the MICE-GC run and compares it to the best-fitting Schechter function found by \citet{Blanton:03} (blue solid line) used in the SHAM technique. The black, blue and red triangles are the luminosity function of total (central+satellite), central and satellite galaxies of the catalogue respectively. The error bars are derived with Poisson statistics. As expected, by following the method to assign luminosities, the agreement between the mock and the data is almost perfect, validating our method to apply the scatter between halo mass and central galaxy luminosity.

\subsection{Galaxy positions}
\label{sec:assign_position}

The standard galaxy formation paradigm assumes that galaxies are formed inside dark-matter haloes, which are overdensities in the underlying dark-matter distribution. Following the HOD ansatz, we populate haloes with central and satellite galaxies following some prescriptions. We then check the clustering properties of the catalogue and iterate on the prescriptions until we get an acceptable fit to observations, using our diagonal-$\chi^2$ (equation~\ref{eq:chi2}). In the halo model formalism the clustering properties are understood in two regimes, one inside haloes and another between haloes. Assuming that dark-matter haloes are small compared to the common separation between them, one can infer that the statistics of the mass density field on small scales are basically governed by the distribution of matter inside haloes (the biggest dark matter haloes have virial radius of $\sim$ 1 Mpc/$h$). Similarly halo mass density profiles are not relevant at scales larger than the size of a typical halo, and therefore the most important contribution on the large scale structure is the spatial distribution of haloes. The two-point correlation function of dark-matter haloes (in the halo model framework) can be divided into two terms:

\begin{equation}
\xi(r)=\xi^{1h}(r)+\xi^{2h}(r)
\end{equation}
The one-halo term takes into account contributions to the density that come from the same halo and the two-halo term takes into account contributions to the density that come from different haloes.

This section describes how we place galaxies inside their host haloes and how we optimize our method to reproduce the observed clustering properties.

\subsubsection{Clustering constraints for centrals}
\label{sec:clustering_constraints_centrals}

As we have indicated before we place central galaxies at the centre of their host haloes and therefore the clustering of central galaxies is the same as the clustering of dark matter haloes. The main goal of our catalogues is to reproduce the spatial clustering as a function of observables like luminosity and colour. In order to characterize the spatial distribution of central galaxies, we compute the two-point correlation function of dark matter haloes as a function of their mass and as a function of the scale. In our method the central galaxy luminosities are completely determined by their halo mass (we will see later that we need to modify this assumption), and therefore the halo clustering for a given halo mass will be equivalent to the central galaxy clustering at a given luminosity. In order to estimate the 2-point correlation function of haloes, we divide the MICE-GC run into $10^{3}$ sub-boxes of size $L_{box}=307.2$ Mpc/h and we compute $\xi_{h}(r)$, from small scales ($\sim0.1$ Mpc/h) up to intermediate scales (10-30 Mpc/h). For computational ease we use the following estimator:

\begin{equation}
\xi_{h}(r) = \frac{DD(r)}{RR(r)} - 1
\end{equation}
where $DD(r)$ refers to the number of pairs in the simulation and $RR(r)$ is the number of pairs as if haloes were spatially randomly distributed. We have checked that results do not change if we use the more common~\citeauthor{Landy:93}  estimator~\citep{Landy:93}, as already discussed in the literature for the scales relevant in this paper~\citep{Kerscher:00}. We compute the two-point correlation function for seven different halo mass thresholds and three halo mass bins. Table \ref{table:Halo_mass_thresholds} shows the chosen halo mass ranges.

\begin{center}
\begin{table}
\scriptsize
\centering
% use packages: array
\begin{tabular}{cccc}
\hline
$\geqslant N_{p}$ & $\geqslant M_{h} (M_{\odot})$& $N_{p1} \geqslant N_{p} > N_{p2}$ & $M_{h1} \geqslant M_{h} > M_{h2} (M_{\odot})$\\
\hline
\hline
10 & 2.2 $\cdot 10^{11}$ & 10 to 39 & 2.2 $\cdot 10^{11}$ to $10^{12}$\\
21 & 5$\cdot 10^{11}$ & 39 to 359 & $10^{12.0}$ to $10^{13.0}$\\
39 & $10^{12}$ & 359 to 3443 & $10^{13.0}$ to $10^{14.0}$\\
179 & 5$\cdot 10^{12}$\\
359 & $10^{13}$\\
1728 & 5$\cdot 10^{13}$\\
3443 & $10^{14}$\\
\hline
\end{tabular}
\caption{\small{Halo mass thresholds and halo mass bins.}}
\label{table:Halo_mass_thresholds}
\end{table}
\end{center}

\begin{figure*}
	\centering
	\includegraphics[width=16cm]{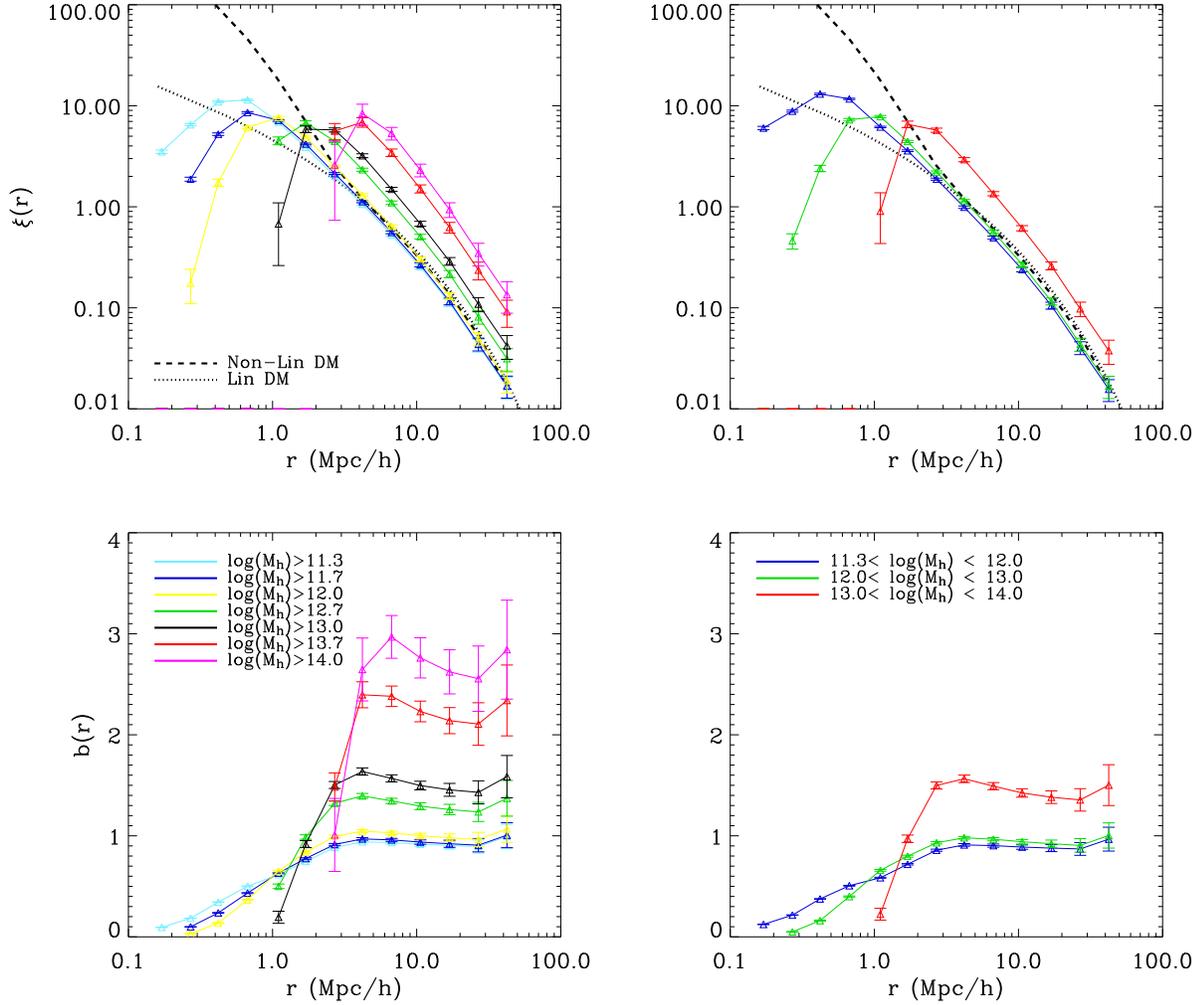}
	\caption{\small{Mean value of the real space 2-point correlation (top panels) and scale dependent linear halo bias (bottom panels). Different colours refer to seven different halo mass thresholds (left panels) and three halo mass bins (right panels). They have been computed using the $10^{3}$ volumes with box-size $L_{box}= 307.2$ Mpc/h. The error bars are the errors on the mean value between volumes. In the two top panels, the dashed black line is the 2-point correlation function of the linear MICE correlation function calculated as the FT of the linear MICE power spectrum, while the black dotted line is the non-linear MICE correlation function.}}
	\label{fig:Halo_CF_Halo_bias_r}
\end{figure*}

The top left and the top right panels of Figure~\ref{fig:Halo_CF_Halo_bias_r} show the average of $\xi_{h}(r)$ for the different halo mass thresholds and bins, respectively. Errors bars are computed taking the error on the mean value between the $10^{3}$ independent sub-volumes. The different colours correspond to different halo mass thresholds and mass bins. We can observe that the three lower halo mass thresholds (up to $\log (M_{h}) \geqslant 12.0 $) have approximately the same amplitude. There are no large differences in the clustering below this halo mass. However, the clustering amplitude increases with halo mass above that halo mass value, $\log (M_{h}) \sim 12.0$.  The errors in the figures are larger at higher halo masses due to the fewer number of haloes. The halo exclusion effect is readily noticeable. There are no halo pairs at short distances. The scale, at which the correlation function turns, shows the approximate size of the haloes. The size of the halo depends on the halo mass. The more massive the halo is, the larger its radius becomes (equation~\ref{r_virial}). For comparison, the linear two-point correlation function of the MICE simulations (calculated as the Fourier Transform of the linear MICE power spectrum) is shown as a black dotted line. The dark matter  two-point correlation function is shown as a black dashed line.

Given the 2-point correlation function, one can estimate the linear large scale halo bias, $b^{Lin}_{h}$, and compare it with theory. We estimate $b^{Lin}_{h}$ using the 2-point correlation functions of the haloes and the dark matter, $\xi_{h}(r)=\left( b^{Lin}_{h}\right)^{2}\xi_{DM}(r)$, assuming a linear bias relation between the haloes and dark matter distributions, $\delta_{h}(\textbf{r}) = b_{h}^{Lin}\delta_{m}(\textbf{r})$. The bottom left and right panels of Figure~\ref{fig:Halo_CF_Halo_bias_r} show the scale dependent halo bias for halo mass thresholds and halo mass bins respectively. The value of the bias is consistent with being constant for scales larger than $\sim$ 5 Mpc/h. The error is larger at large scales because the fewer number of pairs at those distances and the size of the sub-boxes used.

\begin{figure}
	\centering
	\includegraphics[width=8cm]{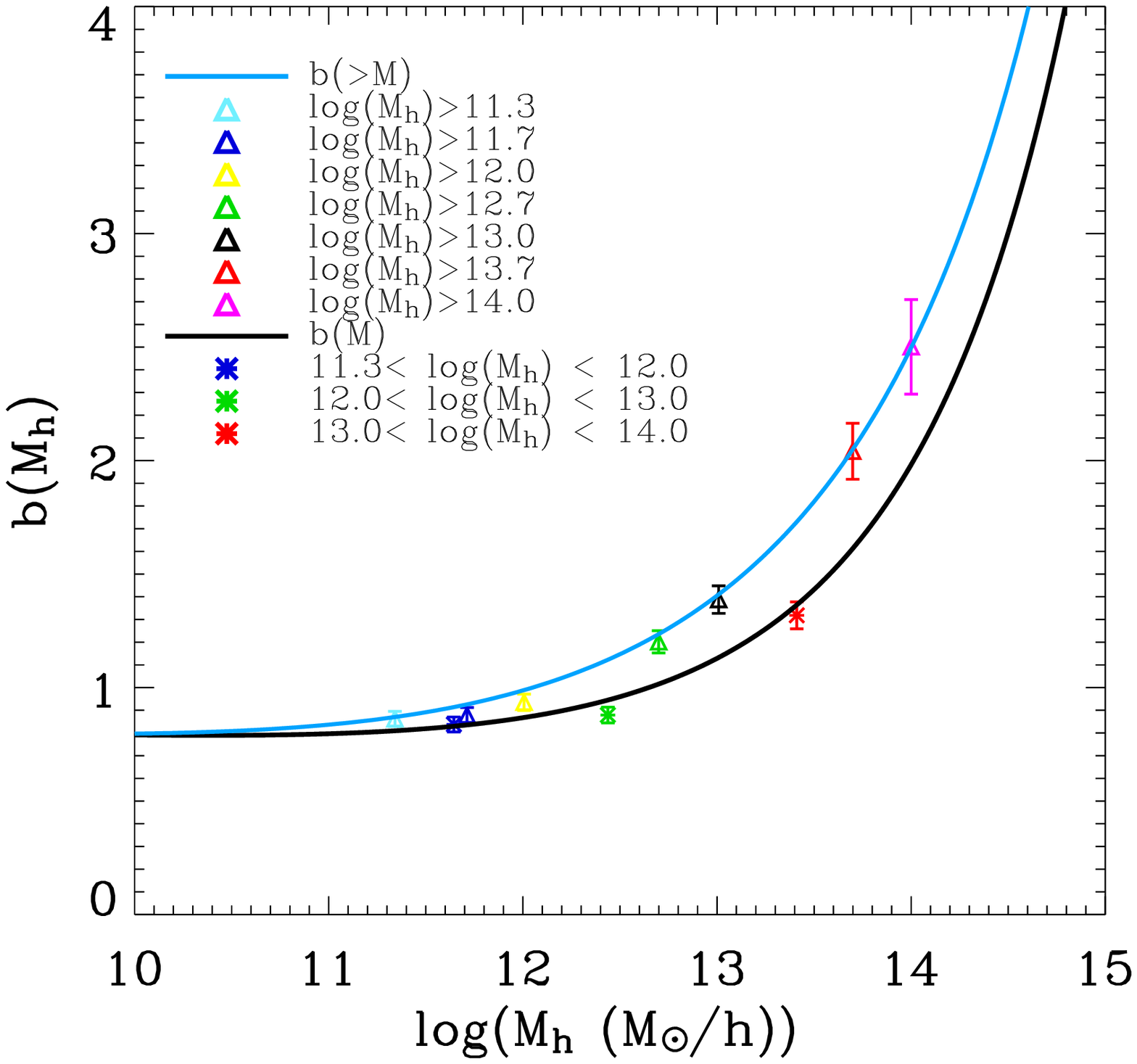}
	\caption{\small{Linear bias at large scales for the halo mass thresholds and halo mass bins using the value of the correlation function at $r = 16.9$ Mpc/h. The black solid line is the expression of the halo bias given by \citet{Manera:10} and using the parameters found by \citet{Crocce:10} for the MICE simulations. The blue solid line is the cumulative bias derived by integrating equation~\ref{eq:Warren bias}}.}
	\label{fig:bias_Halo_Mass_r_16.9}
\end{figure}

In order to better visualize the bias dependence of the central galaxies, Figure~\ref{fig:bias_Halo_Mass_r_16.9} shows the large scale halo bias as a function of mass for different halo mass thresholds and bins. In this case we use the value of the correlation function at $r = 16.9$ Mpc/h, but the figure is very similar if you use also the adjacent radial bins (scales $r=10.6$ or $r = 26.8$ Mpc/h). We compare the MICE-GC simulation halo bias to the fits computed by \citet{Manera:10}, who presented a linear bias function for haloes applying the peak background split to a Warren-form halo mass function (see Appendix~\ref{app:appendix1} for a detailed description of how we have made the comparison). The black solid line is the expression of the halo bias given by~\cite{Manera:10} and using the parameters in~\cite{Crocce:10} for the MICE simulations (see Appendix A for more details). The blue solid line is the cumulative bias derived by integrating equation~\ref{eq:Warren bias}. The MICE-GC large scale halo bias and the analytical expression from~\cite{Manera:10} are in reasonable agreement (for a more detailed study of the clustering of the MICE-GC halo population see~\cite{Crocce:13}).
The shape of the clustering of the central galaxies and its dependence on halo mass (Figure~\ref{fig:bias_Halo_Mass_r_16.9}) will be important to help us build our mock catalogues. 

\subsubsection{Galaxies inside their haloes}
\label{sec:galaxy_position}

We start placing central galaxies at the centre of their host dark matter haloes. Their clustering properties are thus the same as those of their parent haloes, which we will review in section~\ref{sec:clustering_constraints_centrals}.  
Once we have assigned position to the central galaxies, we then turn to the satellite galaxies.  

In the language of the halo model, central galaxies contribute to the two-halo clustering term while the location of satellite galaxies affects both the one-halo and two-halo galaxy clustering regimes. In particular, the satellite contribution is specially relevant for the one-halo clustering term at scales equal or smaller than the size of the haloes, where the non-linear clustering contribution is dominant. 

We then distribute satellite galaxies in their haloes following an NFW mass density profile. However, as discussed below in subsection~\ref{sec:fine_tuning}, we find some tension between the luminosity dependence of galaxy clustering in our mocks compared to observations and thus we have to modify the recipe to place satellite galaxies. This may reveal that our HOD (+ SHAM) modelling is not accurate enough as we use fewer parameters compared to other recent work (e.g., \citealt{Leauthaud:11} use 11 parameters), or that satellite galaxies may not necessarily trace the underlying dark matter distribution. 

High-resolution N-body simulations \citep{Navarro:96, Navarro:97} show that the spherically averaged density around the centres of virialized dark-matter haloes containing mass $M_{h}$ with virial radius $r_{vir}$ can be well by fitted by:

\begin{equation}\label{nfw_profile}
\frac{\rho(r)}{\bar\rho}=\frac{\varDelta_{vir}(z)}{3\Omega(z)}\frac{c^{3}f(c)}{x(1+x)^{2}}
\end{equation}
where $x\equiv c(M_h) \frac{r}{r_{vir}}$, $\bar\rho$ is the average density of the background universe at redshift $z$, $\bar\rho(z)$ $\Delta_{vir}(z)/\Omega_{z}$ is the average density within the virial radius, $f(c)=[\ln{(1+c)}-c/(1+c)]^{-1}$ and $\Delta_{vir}(z)$:

\begin{equation}\label{delta_vir}
\Delta_{vir}(z)=18\pi^{2}+82d-39d^{2}; \quad d=\Omega(z)-1.
\end{equation}
The parameter $c$ is often called the central concentration of the halo. Many authors \citep[e.g.,][]{Bullock:01,Neto:07,Maccio:08,Zhao:09,Munoz-Cuartas:11} have studied the relation between the concentration parameter and the halo mass. As a general trend, more massive haloes are less centrally concentrated. 

We place satellites inside their haloes following equation~\ref{nfw_profile}. In order to do that, besides the cosmological quantities, we need to compute the concentration parameter and the virial radius of the halo to fully define the profile.  We start from the halo mass. We use the parametrization provided by \citet{Bullock:01} to obtain the concentration from the halo mass. 

\begin{equation}\label{concentration}
c(M_{h})\approx\frac{9}{(1+z)}\bigg(\frac{M_h}{M_{\ast_{0}}}\bigg)^{-0.13},
\end{equation}
where $M_{\ast_{0}}$ is the standard non-linear mass scale: $\sigma(M_{\ast})\equiv\delta_{sc}(z)$. We assume $M_{\ast_{0}}=2\cdot10^{13} M_{\odot}$ for $z=0.1$. As we will see later, the exact normalization of the relation turns out not to be important as we will need to modify it when fitting observations. We do not modify the redshift and mass dependence of the relation, though.

We compute the virial radius as:

\begin{equation}\label{r_virial}
r_{vir}=\bigg(\frac{3M}{4\pi\Delta_{vir}(z)\rho_{crit.}}\bigg)^{(1/3)}
\end{equation}

Once the mass density profile is fully defined from the halo mass and the background cosmology, we need to specify how to place the satellites inside the haloes. Given the number of satellites in each halo extracted form the previous realization of the HOD parametrization (see section~\ref{sec:assign_position}), we Montecarlo sample the density distribution to obtain the radial position of the individual satellite galaxies. We leave as a free parameter the outermost radius until which we place satellite galaxies. As a first approximation, we assume that the haloes are spherical and draw randomly two position angles that maintain uniform the surface density in a sphere. We later modify this assumption and make our haloes triaxial (section~\ref{sec:fine_tuning}). Once we have the positions of the central and satellite galaxies, we compute the clustering properties of the catalogue and compare them to observational constraints to check whether our recipes reproduce the observed properties. 

%Placing satellite galaxies following an NFW mass density profile with the concentration index given by equation~\ref{concentration} makes the galaxy clustering at very small scales (from $\sim0.1$ to $\sim0.5$ Mpc/h) have a smaller amplitude than observations. 
%Subsection~\ref{sec:fine_tuning} explains the fine tuning that we have introduced when placing satellite galaxies inside dark matter haloes to better match observations.

\subsubsection{Clustering constraints from satellite galaxies}
\label{sec:fine_tuning}

We now turn our attention to how the satellite luminosities and positions affect the overall clustering. As described previously, we assign luminosities to satellite galaxies drawing them from the global satellite cumulative luminosity function (equation~\ref{eq:nsatL}). We assign their positions assuming that they are distributed following an NFW profile with the concentration index given by equation~\ref{concentration} (see section~\ref{sec:galaxy_position}). Satellite galaxies define the one-halo clustering term and provide a smaller contribution to the clustering amplitude than central galaxies in the two-halo term regime as there are always fewer number of satellites than centrals at the luminosities considered in this work (as we will see later in section~\ref{sec:assign_colour} and Figure~\ref{fig:fraction_objects}).    

When we compute the projected galaxy correlation function, $w_{p}(r_{p})$, following the previous recipes for satellites, we find that we are unable to fit observations. Our clustering amplitude is lower than observed at small scales, $r_p \lesssim  1.0$ Mpc/h (this value depends on luminosity as in the top panels of Figure~\ref{fig:Halo_CF_Halo_bias_r}), corresponding to the one-halo term range. This lower amplitude at small scales is present at all luminosities. We find that to increase the clustering amplitude at these scales we need to concentrate satellite galaxies more than the NFW dark matter distribution provided by equations~\ref{nfw_profile}, \ref{delta_vir}, \ref{concentration} and~\ref{r_virial}. 
\cite{Watson:12} find the same trend with similar assumptions to ours. Satellite galaxies need to be more concentrated than their underlying dark matter NFW density profile.

In order to place satellites closer to the cluster centre, we apply a factor, $f$, to the previously computed distance for each galaxy sampling the NFW profile. As the amount of mismatch between the computed $w_{p}(r_{p})$ and the observed one is different for different luminosity samples, we make the $f$ correction factor depend on luminosity to stay closer to the observational data.
For each luminosity bin sample, we compute the $f$ factor that better corrects the projected correlation function
to coincide with observation. Once we have those values for the different luminosity samples, we combine them with the following expression, where we have two values at bright and faint magnitudes and a linear interpolation in between:

\begin{equation}\label{sat_concentration}
	\begin{array}{ll}
		f=0.30 \ \ \ \ \ \ \ \ \ \ \ \ \ \ \ \ \ \ \ \ \ \ \ \ \ \ \ \ \ \ \ \ \ \ \ \ \textrm{ if } M_{r} > -19.25 \\
		f=0.30+\frac{0.65-0.30}{-21.0+19.25}(M_{r}+19.25) \textrm{ if } -21 \leqslant M_{r} \leqslant -19.25Ê\\
		f=0.65 \ \ \ \ \ \ \ \ \ \ \ \ \ \ \ \ \ \ \ \ \ \ \ \ \ \ \ \ \ \ \ \ \ \ \ \ \textrm{ if } M_{r}  < -21.0		
	\end{array}
\end{equation}

This correction changes the clustering properties in the one-halo term regime as it only influences how satellites are placed inside haloes. The correction increases the clustering amplitude as we concentrate more the satellite galaxies than the expected NFW profile given the halo mass. 

Satellite galaxies also contribute to the clustering two-halo term beyond the typical virial radius of haloes at scales $r_p\gtrsim 1 Mpc/h$. The transition scale as a function of halo mass is clearly distinguishable in both top panels of Figure~\ref{fig:Halo_CF_Halo_bias_r}. As we will see later, satellite galaxies constitute approximately 30\% of the galaxy population at absolute magnitudes fainter than $M^{*}_{r}$ and progressively a smaller percentage for brighter magnitudes (Figure~\ref{fig:fraction_objects}). 
Their contribution to the overall amplitude of the two-point correlation is therefore smaller than the one due to central galaxies especially at bright absolute magnitudes. As we have seen before, at the high luminosity end we had to introduce scatter in the halo mass-luminosity relation to place central in lower mass haloes to reduce the clustering amplitude. The contribution of satellites is much smaller at these high luminosities and the way we distribute them in their haloes is almost irrelevant to the resulting large scale clustering. The important parameter at these scales and luminosity ranges is the occupation distribution provided by the HOD. At fainter magnitudes, the situation is different. The halo bias is flat with respect to halo mass and therefore the clustering of centrals cannot practically be changed introducing scatter in the halo mass-luminosity relation dictated by abundance matching. Changes in the overall clustering amplitude are then dominated by what the satellite distribution is. The HOD sets the number of satellites per halo as a function of halo mass and the abundance matching sets the luminosity distribution of the overall satellite population. We have chosen to assign those luminosities to individual satellites in a random fashion (see section~\ref{sec:galaxy_luminosity}). The clustering of the satellite galaxies depends on the HOD parametrization but also on how the luminosities are assigned to individual satellites in our method. While the HOD parameters will affect the overall clustering strength of the satellite galaxies, 
with the luminosity assignation we can modify the two-halo term clustering of the satellites without noticeably altering the one-halo term. 
In particular, in the implementation of the method to produce the MICECAT v1.0 catalogue presented in this paper, we find that the bias is still somewhat stronger than observed at the low luminosity galaxy bins. We therefore choose to penalize the assignation of low luminosity satellites to the most massive haloes. In this way, as low luminosity satellites are now in slightly less massive haloes, their clustering amplitude is lowered. This comes at a price of slightly increasing the bias at brighter luminosities. Nevertheless, with a judicious penalization algorithm, one can slightly transfer clustering power between satellite luminosity bins. In our case, we arbitrarily set a line in the luminosity-halo mass plane and set a probability of rejecting the assignation of a luminosity to a satellite in a particular halo based on the distance to this line\footnote{In detail, we set a straight line in the luminosity-halo mass plane, $\log{L_{sat}}=a_{0}\log{M_{h}}+b$, where $a_{0}=0.36$, $b=4.86$. Then for each satellite galaxy that inhabits in a given halo mass, we derive the distance to the straight line from the point formed by the halo mass and one of the $N_{sat}$ sampled satellite luminosities. We draw a uniformly distributed random number, $u_{3}$, between 0 and 1. Then we assign one of the sampled satellite luminosities if $d < 0$ (the sampled satellite luminosity is not in the region corresponding to faint luminosity and massive halo) or $u_{3}>1-\exp{\left(-7d\right)}$ (the distance to the straight line is not very large).}.
This way we manage to reduce the clustering amplitude at low luminosities to better match observations.

\begin{center}
\begin{table}
\centering
% use packages: array
\begin{tabular}{cc}
\hline
$< M_{r}$ & $M_{r1} < M_{r} < M_{r2}$\\
\hline
\hline
-19.0 & -19.0 to -20.0\\
-19.5 & -20.0 to -21.0\\
-20.0 & -21.0 to -22.0\\
-20.5 & -22.0 to -23.0\\
-21.0\\
-21.5\\
-22.0\\
\hline
\end{tabular}
\caption{Absolute magnitude thresholds and bins in the r-band.}
\label{table:Lum_th_bins}
\end{table}
\end{center}

Numerical N-body simulations find that dark mater haloes are not spherical and can be better represented as triaxial systems~\citep[e.g.,][]{Jing:02}.
\citet{vanDaalen:12} have studied the effect of halo shape on the clustering of haloes. They find that the clustering amplitude is lower at the one-halo term regime when haloes shapes are made spherical. Taking into account these results and as a way to populate haloes in a more realistic fashion, we have placed satellite galaxies following a triaxial distribution. For the catalogue version we have released we simply assume that all haloes are triaxial with axis ratios q=b/a=0.8 and s=a/c=0.6, following the most representative values presented in other work~\citep[e.g.,][]{Jing:02}. Contrary to~\citet{vanDaalen:12}, we find that the effect on the galaxy clustering is minor compared to assuming spherical haloes but that may be just a result of the orientation of the haloes that is random in our case without following the tidal forces of large scale structure. A detail study of these effect is beyond the scope of this work and we defer a proper treatment of the effects of triaxiality to further work.

\subsubsection{Clustering as a function of luminosity}
\label{sec:clustering_luminosity}

\begin{figure*}
	\centering
	\includegraphics[width=16cm]{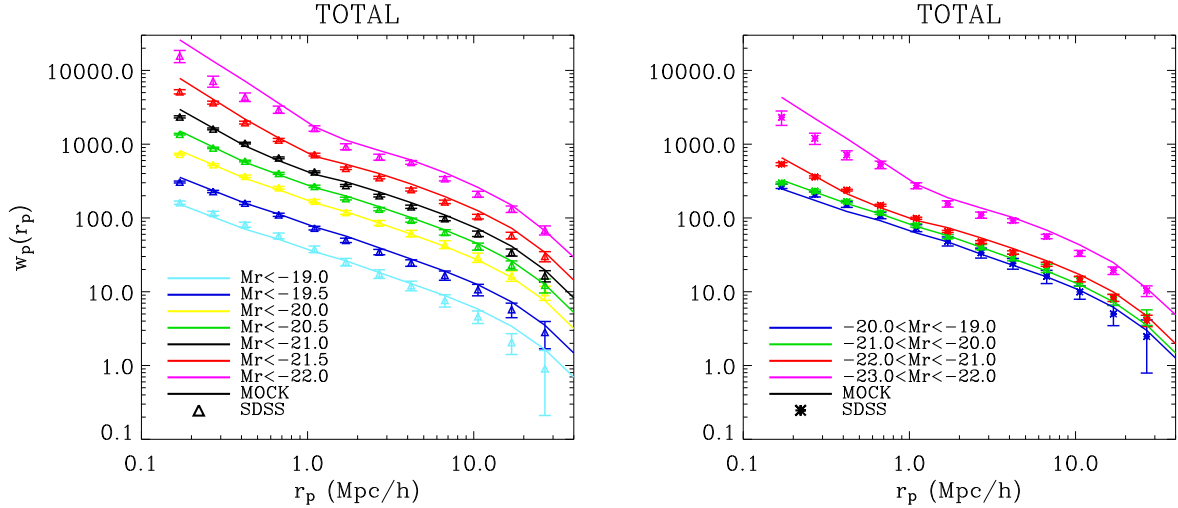}
	\caption{\small{Luminosity dependence of galaxy clustering. Data points come from SDSS data (see section \ref{sec:intro} for more details) and solid (coloured lines) refer to different luminosity threshold (left panel) and bin samples (right panel) from the galaxy catalogue. In the left panel, the samples are staggered for clarity since galaxies fainter than $M^{*}_{r}$ have almost the similar bias.}}
	\label{fig:wp_th_bins_plot}
\end{figure*}

\begin{figure*}
	\centering
	\includegraphics[width=16cm]{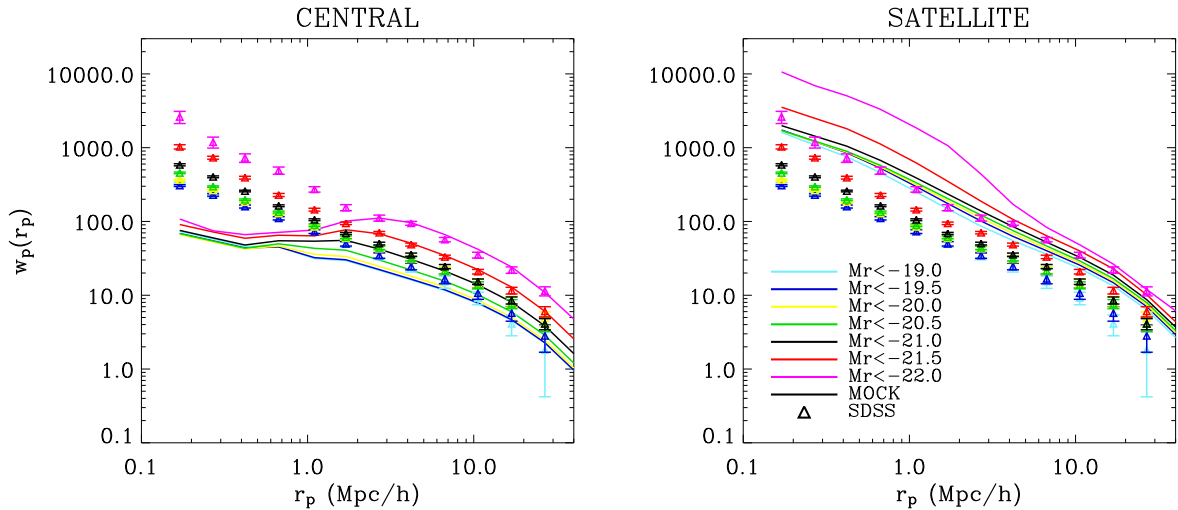}
	\caption{\small{Luminosity dependence of central (left panel) and satellite (right panel) galaxy clustering. Figure~shows the projected correlation functions for different luminosity threshold samples of SDSS and (central and satellite) galaxies of the mock galaxy catalogues. In this case samples are not staggered. Data points come from SDSS data (see section \ref{sec:intro} for more details) and solid (coloured lines) refer to different luminosity threshold samples of the mock galaxy catalogue.}}
	\label{fig:wp_th_bins_plot_cen_sat}
\end{figure*}

We estimate the two-point projected correlation function, $\omega_{p}(r_{p})$, for seven different luminosity thresholds and four luminosity bins subsamples (see table \ref{table:Lum_th_bins}). These subsamples correspond to the same luminosity subsamples observed in the SDSS and presented in Z11. The left panel of Figure~\ref{fig:wp_th_bins_plot} shows the luminosity dependence of galaxy clustering for the different luminosity threshold subsamples produced by our method compared to the SDSS data. The values of the correlation function of the subsamples include an offset to stagger them for clarity since galaxies fainter than $M^{*}_{r}$ have almost the same bias. 
Data points come from SDSS data (see section \ref{sec:input} for more details) and solid lines correspond to the generated mock galaxy catalogue. We have used the comparison between the two to optimize and tweak our method. The agreement is reasonably good, with values of our diagonal-$\chi^2/d.o.f < 1$, although at large scales the amplitude of the galaxy clustering of the galaxy catalogue is in general slightly larger than that of the SDSS data, specially for galaxies brighter than $M^{*}_{r}$. However, as explained previously, if we do not include scatter in the relation $M_{halo}-L_{gal}$ the amplitude of the clustering at large scales would be much higher. 
The right panel of Figure~\ref{fig:wp_th_bins_plot} shows the projected correlation function of luminosity-bin samples of the catalogue compared to SDSS data (Z11). Again, data points come from SDSS data and the solid lines correspond to the mock galaxy catalogue. The fits are quite reasonable with acceptable values of the $\chi^2$. 

In Figure~\ref{fig:wp_th_bins_plot_cen_sat}, we show separately the projected two-point correlation function of central (left) and satellite (right) galaxies. At the same scales, the clustering of central galaxies is lower than that of satellites. At the two-halo term scales, the clustering of centrals dominates the overall clustering (Figure~\ref{fig:wp_th_bins_plot}, left panel), while satellites are the contributors to the clustering at the one-halo scales.

\begin{figure*}
	\centering
	\includegraphics[width=16cm]{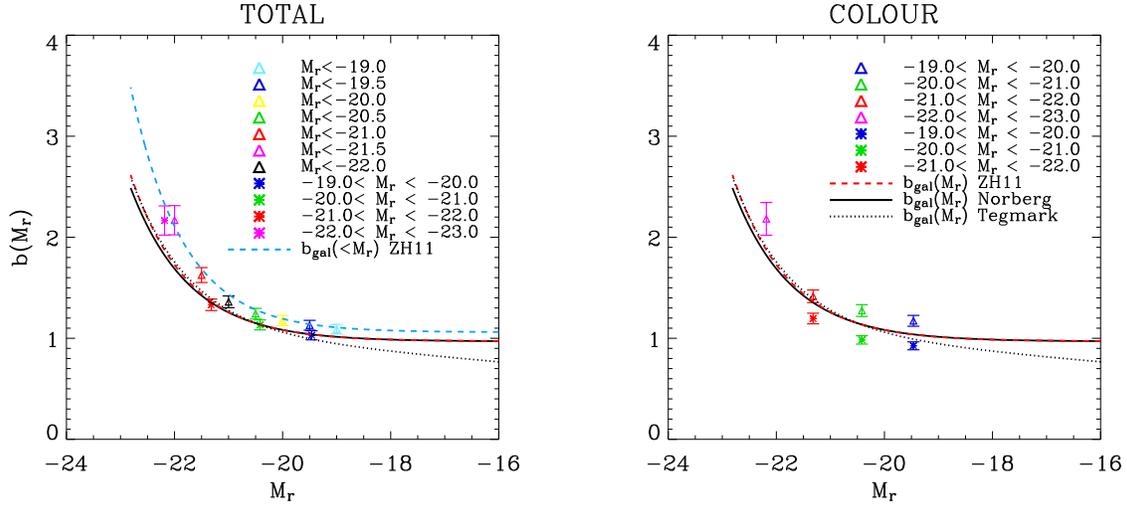}
	\caption{\small{Galaxy bias as a function of $M_{r}$ of luminosity threshold, $b_{g}(<M_{r})$, and bin samples, $b_{g}(M_{r})$,  for the whole catalogue (left panel), and $b_{g}(M_{r})$ for red and blue galaxies (right panel). Left panel: the coloured triangles and asterisks represent luminosity threshold and bin samples respectively. The light blue dashed line is the fit to the HOD model bias factors for luminosity threshold samples from~\citet{Zehavi:11} (equation~\ref{eq:bias_th_ZH11}). The red dashed line is the fit to the HOD model bias factors for luminosity bin samples from \citealt{Zehavi:11} (equation~\ref{eq:bias_bins_ZH11}). The black solid line is a fit to projected correlation functions in the 2dFGRS (equation~\ref{eq:bias_bins_Norberg}) and the dashed curve is a modified fit to SDSS power spectrum measurements derived by~\citet{Tegmark:04} (equation~\ref{eq:bias_bins_Tegmark}). Right panel: the coloured triangles correspond to red luminosity bin samples and the coloured asterisks correspond to blue luminosity bin samples. The lines are the same as in the left panel for the luminosity bin samples.}}
	\label{fig:gal_mean_bias_Mr_CF_3D_total_colour}
\end{figure*}

\begin{figure*}
	\centering
	\includegraphics[width=16cm]{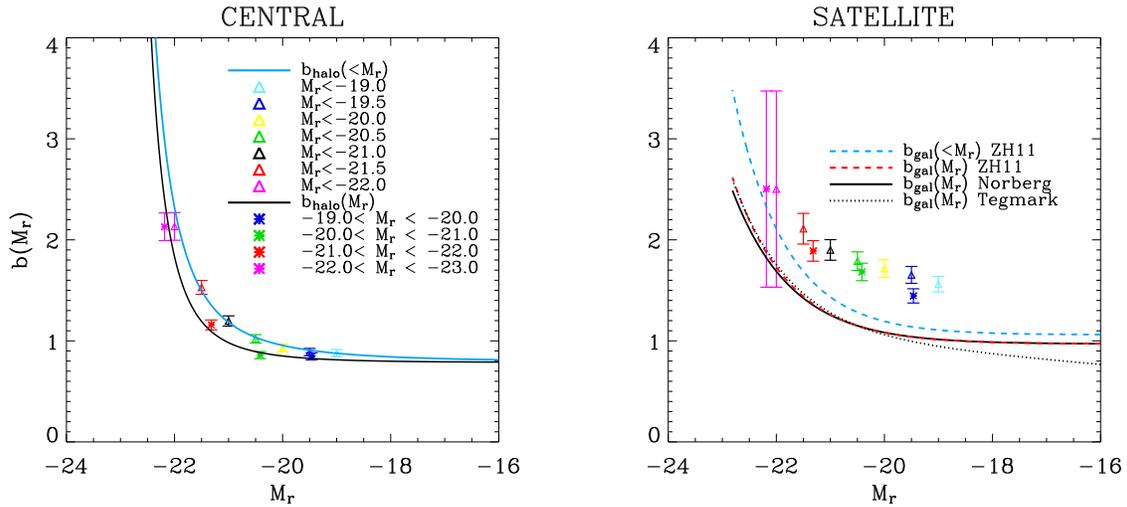}
	\caption{\small{Galaxy bias as a function of $M_{r}$ for central (left panel) and satellite galaxies (right panel). For both panels, the coloured triangles and asterisks represent luminosity threshold and bin samples respectively. Left panel: the black line is the halo bias derived by~\citet{Crocce:10} for the MICE simulations relating halo masses to absolute magnitudes by using the relation $M_{h}-L_{gal}$ derived using the subhalo abundance matching technique. The blue line is derived by integrating the black line. Right panel: the lines are the same as the ones plotted in Figure~\ref{fig:gal_mean_bias_Mr_CF_3D_total_colour}.}}
	\label{fig:gal_mean_bias_Mr_CF_3D_cen_sat}
\end{figure*}

Finally, in Figure~\ref{fig:gal_mean_bias_Mr_CF_3D_total_colour} we show the linear large scale galaxy bias as a function of $M_{r}$. In this case, after calibrating the method, we apply it using the whole snapshot at $z=0$ of the MICE-GC run. We again divide the whole volume into $10^{3}$ sub-boxes and compute the two-point correlation function for each sub-volume. The value of the bias we use in the following plots and discussion is the average of the bias at scales $r_{1}=10.6$ and $r_{2}=16.9$ Mpc/h\footnote{These scales come from the binning chosen by Z11.}. At these scales the bias is fairly flat as seen in the bottom panels of Figure~\ref{fig:Halo_CF_Halo_bias_r}. Therefore the exact scale at which we measure it is somewhat irrelevant and can be considered representative of these range of scales. The left panel of Figure~\ref{fig:gal_mean_bias_Mr_CF_3D_total_colour} shows $b_{g}(<M_{r})$ and $b_{g}(M_{r})$ for the whole catalogue. The triangles refer to luminosity threshold samples and the asterisks represent luminosity bin samples. The blue dashed line is a fit to the HOD model bias factors derived by Z11 given by the following expression:

\begin{equation}\label{eq:bias_th_ZH11}
b_{g}(>L_{r})\cdot(\sigma_{8}/0.8) = 1.06 + 0.21(L_{r}/L^{*}_{r})^{1.12}
\end{equation}
where $L_{r}$ is the r-band luminosity corrected to $z = 0.1$ and $L_{r}^{*}$ corresponds to $M_{r}^{*} =-20.44$ (\citealt{Blanton:03}). The red dashed line is the fit of Z11 to the luminosity bin samples and is given by the following expression:
\begin{equation}\label{eq:bias_bins_ZH11}
b_{g}(L_{r})\cdot(\sigma_{8}/0.8) = 0.97 + 0.17(L_{r}/L^{*}_{r})^{1.04}
\end{equation}
We have also plotted, as is done in Z11, the formula derived by \citet{Norberg:01} to fit the projected correlation functions in the 2dFGRS for $b_{J}-$selected galaxies (black solid line):

\begin{equation}\label{eq:bias_bins_Norberg}
b_{g}/b^{*} = 0.85 + 0.15(L/L^{*})
\end{equation}
where we take $b^{*} \equiv b_{g}(L_{r}^{*})=1.14$ as in Z11, and a modified fit to SDSS power spectrum measurements derived by~\citet{Tegmark:04} (dotted black line):

\begin{equation}\label{eq:bias_bins_Tegmark}
b_{g}(>L_{r})/b^{*} = 0.85 + 0.15(L_{r}/L^{*}_{r})-0.04(M_{r}-M^{*}_{r})
\end{equation}
The agreement between the bias of luminosity threshold samples of the catalogue with the fit of Z11 is quite reasonable, which is not unexpected given that the mock galaxy catalogue is constructed to match the SDSS data presented in their work. The agreement between the bias of luminosity bin samples with the fit obtained by Z11 is also very good except for the brightest samples where our catalogue is more clustered than the fit, although at these luminosities the errors are larger. At fainter magnitudes the agreement with the fit of \citet{Norberg:01} is remarkably good and also with the fit of \citet{Tegmark:04}.

In Figure~\ref{fig:gal_mean_bias_Mr_CF_3D_cen_sat} we show the bias as a function of $M_{r}$ for central (left panel) and satellite galaxies (right panel). In both panels, the triangles refer to luminosity threshold samples and the asterisks refer to luminosity bin samples. In the left panel, the blue and black solid lines are derived using the analytical expression for the MICE halo bias (combining the expression of \citet{Manera:10} and the parameters found by \citet{Crocce:10} for the halo mass function) and converting halo masses to luminosities by using the relation, $M_{halo}-L_{gal}$, obtained when building the mock galaxy catalogue. The agreement is very good for both, luminosity threshold and bin samples. The bias of central galaxies should be similar to the results shown in Figure~\ref{fig:bias_Halo_Mass_r_16.9} for haloes since they are placed at the centre of their host haloes. Now we plot against luminosity while before we were plotting against halo mass. The scatter introduced in the relation $M_{halo}-L_{gal}$ when assigning central luminosities makes the values of the bias to be somewhat different at the bright and massive end. In the right panel we have also plotted the fits derived by Z11, \citet{Norberg:01} and \citet{Tegmark:04} for comparison. As explained before, satellite galaxies, at the same luminosity, are much more clustered than central galaxies (Figures~\ref{fig:wp_th_bins_plot_cen_sat} and \ref{fig:gal_mean_bias_Mr_CF_3D_cen_sat}). The bias variation with luminosity is similar to that of haloes, it is almost flat for galaxies fainter than $M_{r}^{*}=-20.44$ and it increases very steeply for brighter luminosities. It is also worth noticing that faint central galaxies are anti-biased, while satellites have always $b_{sat}(L_{r})>1$ in the luminosity range covered by our catalogue. Overall, the whole population at faint luminosities is almost unbiased. A more detailed analysis of the clustering of the catalogue can be found in~\citet{Crocce:13}.

\begin{figure}
	\centering
	\includegraphics[width=8cm]{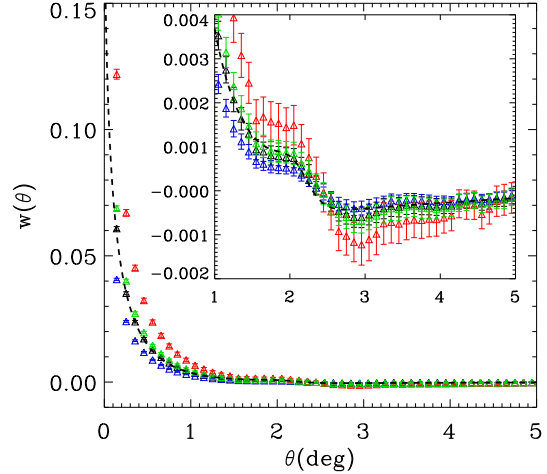}
	\caption{\small{Angular correlation function at $z\sim1.35$ for total (black), central (blue) and satellite (red) galaxies. The green triangles refer to the cross-correlation between central and satellite galaxies. The black dashed line is the theoretical angular correlation function of MICE simulations. The inset is a zoomed view.}}
	\label{fig:w_theta_multi}
\end{figure}

So far we have focused on the clustering at small scales as those are the most constraining for calibrating our method. 
Note that our catalogues are also consistent with the clustering expected at larger scales, as has been already extensively discussed in~\citet{Crocce:13}. Complementing that study, we now show in Figure~\ref{fig:w_theta_multi} the angular two-point correlation function, $w({\theta})$, at large scales including the Baryonic Acoustic Oscillations (BAO) scale. In particular we derive $w(\theta)$ for a comoving-distance spherical shell of width $300$ Mpc/h, covering the range $(2772-3072)$ Mpc/h. The mean value of the shell is $z\sim1.35$ in the MICE cosmology.

Figure~\ref{fig:w_theta_multi} shows the $w(\theta)$ for the total (black), central (blue) and satellite (red) galaxies. The cross-correlation between central and satellite galaxies (green) is also shown. The black dashed line is the theoretical angular correlation function of the MICE simulations. 
%In order to compute it we use the MICE power spectrum, then we compute its Fourier transform to get the two point correlation function (we normalize it to get $\sigma_{8}=0.8$). Finally, we integrate to obtain the projected correlation and transforming the distance into angles using the mean value of the redshift bin. 
The inset shows an expanded view of Figure~\ref{fig:w_theta_multi} in order to better show the region around the acoustic peak 
%and where the correlation function becomes negative. Looking at the whole figure one can observe that the angular correlation function decreases very fast at small angles until $\theta\sim0.5$ and then becomes almost flat around zero. Looking at the inset, $w(\theta)$ is decreasing until $\theta\sim1.5$ where the amplitude starts to be almost flat at $\sim0.5$ deg, in this angle interval is where the $\theta_{BAO}$ should be, then decreases and becomes negative ($\theta\sim2.3$), and then increases slowly towards zero. 
As described in previous sections, one can observe that, at the same luminosity, satellite galaxies (red) are the most clustered. It is interesting to notice that the cross-correlation between central and satellite galaxies is less clustered than satellite galaxies themselves but it is more clustered than the autocorrelation of the whole catalogue and the central catalogue. 
%Looking for example at the satellite clustering, another aspect to notice, is that the amplitude of the clustering when the correlation is positive is the largest, however it is the smallest when the correlation is negative. This behaviour is also expected since the integral of the correlation function over the whole angle range is zero. The error-bars are larger for satellite galaxies because the number of satellites is smaller than the number of objects in the other galaxy samples.

\subsection{Galaxy velocities}

This section describes how we assign velocities to the galaxies in the catalogue. We also qualitatively show what is the effect of including galaxy velocities in the clustering properties of the catalogue. A detailed study on redshift space distortions in our mock galaxy catalogue is presented in~\citet{Crocce:13}. Their analysis shows that the distortions in the clustering induced when assigning velocities to our galaxies are consistent with theoretical expectations.

There are basically two common different approaches to assign galaxy velocities when populating N-body simulations with galaxies. One option is to directly assign the dark matter particles velocities to the galaxies. The other consists on following a model that describes the galaxy velocity distribution. As it was shown in section \ref{sec:assign_position}, we have not assigned galaxies to dark matter particles that form the dark matter halo in which galaxies inhabit. If this were the case, assigning dark matter velocities to galaxies would be the logic approach. In our case, since we have placed galaxies inside their host halo following some mass density profile, we also model the galaxy velocity distribution with simplistic assumptions.

The motion of a particle in a N-body simulation can be described as the sum of the virial motion of the particle within its host halo and the bulk motion of the halo as a whole (\citealt{Sheth:01}),

\begin{equation}
v=v_{vir}+v_h
\end{equation}
The virial motions within a relaxed halo are well approximated by a Gaussian distribution with velocity dispersion that depends on halo mass as: $\sigma^{2}_{vir}=\,<v^{2}_{vir}>\, \propto GM_h/r_{vir} \propto M^{2/3}_h$. To obtain the velocity dispersion of the halo we use (\citealt{Bryan:98}):

\begin{equation}\label{eq:sigma_vir}
\sigma_{vir}=476f_{vir}[\Delta_{vir}E^2(z)]^{1/6}\bigg(\frac{M_h}{10^{15}M_{\odot}h^{-1}}\bigg)^{1/3} km s^{-1},
\end{equation}
where $f_{vir}$=0.9 and $E^{2}(z)=\Omega_{0}(1+z)^{3}+\Omega_{\Lambda}$ for a flat model with a cosmological constant. We prefer to use an analytical expression than directly derive the velocity dispersion from the motions of the dark matter particles in the halo. In order to obtain catalogues complete down to faint magnitudes we have pushed the selection of haloes down to friend-of-friends associations of ten dark matter particles. Sampling the velocity distribution with so few particles can lead to errors and therefore we prefer to use the halo mass as the property that defines the dynamics of our systems. This provides also self-consistency as we have also used the halo mass for defining the HOD and SHAM implementations. 

In our catalogue, central galaxies are assumed to be at rest at the centre of the halo with no peculiar velocity. Satellite galaxies are assigned a velocity vector that is drawn from the velocity dispersion of the halo. We assume that the projection of the velocity vectors on the three cartesian axes have independent Gaussian distributions with dispersions a factor $1/\sqrt{3}$ of the global halo velocity dispersion value. 

\begin{equation}
\sigma_{vx} = \sigma_{vy} = \sigma_{vz} = \frac{\sigma_{vir} }{\sqrt{3}}
\end{equation}
We therefore draw three independent realizations of that distribution to assign each Cartesian component of the satellite velocity vector. Velocities are thus assumed to be independent of the satellite position in the halo.
The overall scheme we have chosen to assign velocities is fairly simplistic. We neglect effects that are important to properly describe the velocity of galaxies within haloes. However, our treatment should be adequate for the dynamics on scales larger than the typical halo size as described in \cite{Crocce:13}. But we warn the reader that they should take into account the limitations of the catalogue for the study of the dynamics within haloes.

The distortions in the galaxy clustering when measuring galaxy redshifts due to their peculiar velocities can be visualized by comparing the two-point correlation function in real and redshift space. The observed redshift is the sum of the redshift due to the expansion of the universe and the redshift due to the peculiar velocity of the galaxy. There are two well-known effects, the so-called Fingers-of-God effect, and the Kaiser effect~\citep{Kaiser:87}.
The dynamics of galaxies within haloes makes the distribution of galaxies in redshift space be elongated along the line of sight (like fingers). For example, galaxies in the same halo and at the same distance but with different velocities are observed at different redshifts. 
The Kaiser effect is evident at larger scales. Galaxies fall into overdense regions making them appear further away if they are in front of the overdensity and closer by if they are behind the overdensity with respect to the observer. The net effect is a squashing in the line of sight of the galaxy distribution in redshift space. We can visualize these two effects when computing the two-point correlation function as a function of radial or line of sight distance, $r_{\pi}$, and the perpendicular distance, $r_{p}$, in the $\xi(r_{p},r_{\pi})$ diagram since both effects are going to produce distortions in the observed galaxy positions and therefore distortions in the number of pairs depending on the distance at any scale.

\begin{figure}
	\centering
	\includegraphics[width=8cm]{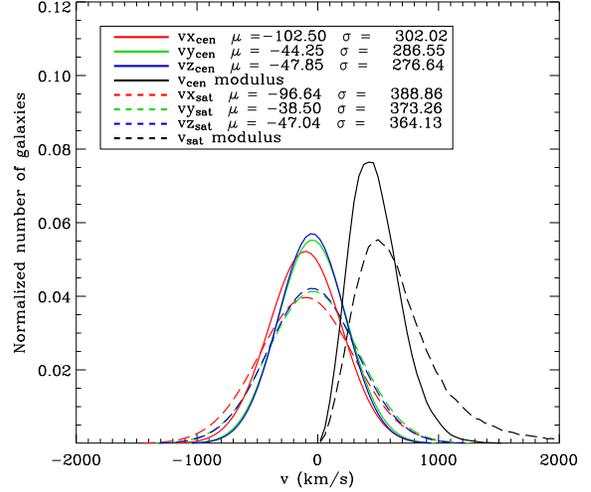}
	\caption{\small{Galaxy velocity histogram. Solid and dashed lines show the Gaussian fits to the three different components of the central and satellite galaxy velocity, respectively. Black lines show the modulus of the velocities.}}
	\label{fig:vgal_norm}
\end{figure}

Figure~\ref{fig:vgal_norm}  shows the distribution of galaxy velocities in a sub-box with size $L_{box}=307.2$ Mpc/h of the snapshot at $z=0$ of the MICE-GC run. The normalized distribution of the three different velocity components are shown as well as the distribution of the velocity modulus for central (solid lines) and satellite galaxies (dashed lines). It is worth noticing that the mean values for all velocity components are non-zero, indicating that even at such very large scales ($\sim L_{box}$) there exist bulk motions. 

Once we know the comoving-distance of a given galaxy to obtain its cosmological redshift we use the relation: 

\begin{equation}
r_{com}(z)=\int_{0}^{z}\frac{c dz'}{H(z')}
\end{equation}
where $c$ is the speed of light and $H(z)$, the Hubble constant at redshift $z$. We compute the observed position  in the radial direction distorted due to the velocity component in the radial direction $v_{||}$ as: 

\begin{equation}\label{eq:RSD_cosmo}
r_{||}^{obs}=r_{||}^{com}+\frac{v_{||}}{aH(z)}
\end{equation}
where $a=\frac{1}{1+z}$ is the scale factor. When studying snapshots we assume that all galaxies in the box are at the same redshift. This is not the case when studying the MICE-GC lightcone as in~\citet{Crocce:13}.

\begin{figure*}
	\centering
	\includegraphics[width=16cm]{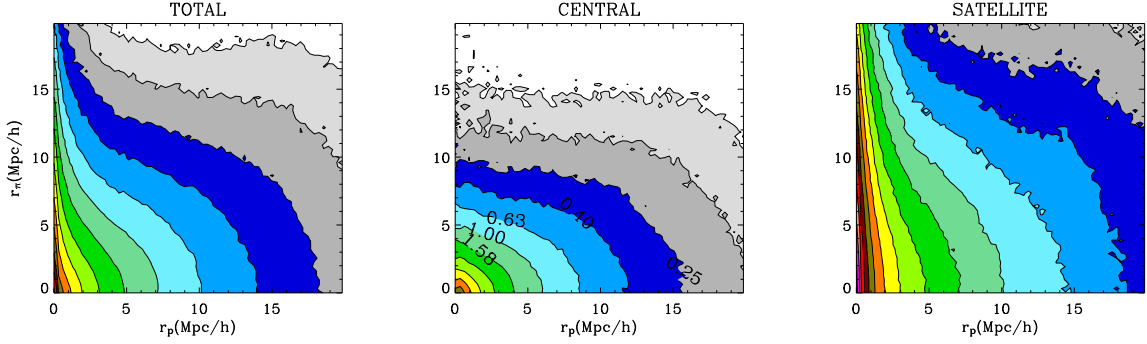}
	\caption{\small{Two-point correlation function $\xi(r_{p},r_{\pi})$ as a function of radial or line of sight ($r_{\pi}$ in vertical axis) and perpendicular distance ($r_{p}$ in horizontal axis) for a sample of galaxies brighter than $M_{r}< -19.0$, in redshift space (taking into account the velocity of galaxies). Left, central and right panels correspond to the total (central+satellite), central and satellite catalogues respectively. The maximum values of the correlation, which correspond to the lowest distance, $(r_{p},r_{\pi})=(0.16,0.16)$ Mpc/h, for total, central and satellite galaxies are $\sim$ 118, 41 and 13 respectively (red and orange colours), while grey and blue colours correspond to small correlation (e.g., the correlation amplitude for dark blue colour is $\sim$ 0.35).}}
	\label{fig:RSD_multi_19_0}
\end{figure*}

\begin{figure*}
	\centering
	\includegraphics[width=16cm]{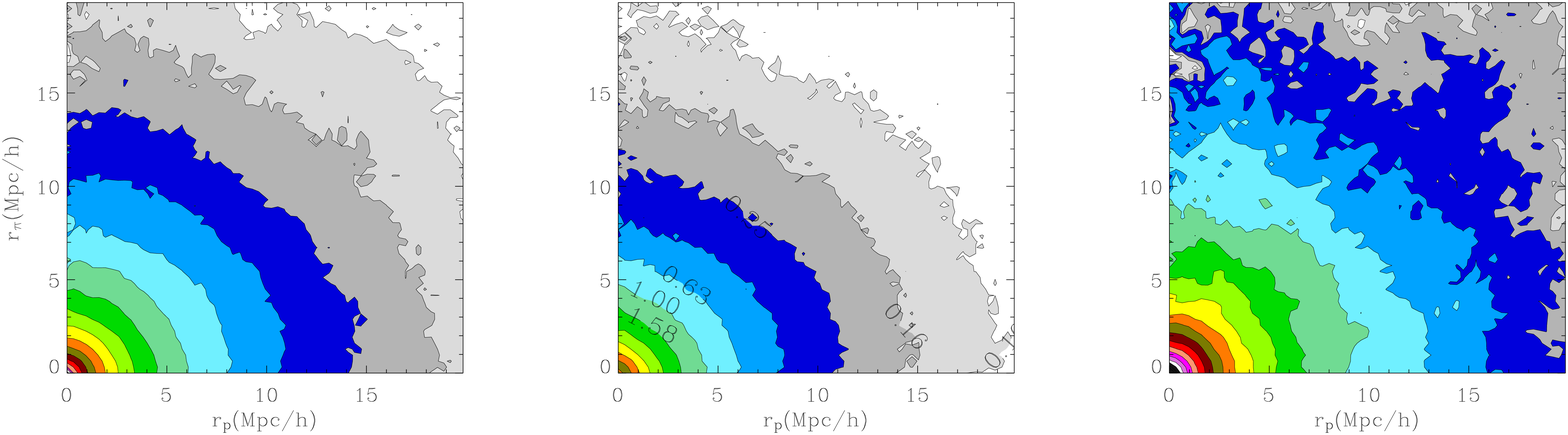}
	\caption{\small{Two-point correlation function $\xi(r_{p},r_{\pi})$ as a function of radial or line of sight ($r_{\pi}$ in vertical axis) and perpendicular distance ($r_{p}$ in horizontal axis) for a sample of galaxies brighter than $M_{r}< -19.0$ in real space (without taking into account the velocity of galaxies). As in Figure~\ref{fig:no_RSD_multi_19_0}, left, central and right panels correspond to the total (central+satellite), central and satellite catalogues respectively. Orange and red colours correspond to large correlation and blue and grey colours to small correlation.}}
	\label{fig:no_RSD_multi_19_0}
\end{figure*}

Here, we present a qualitative description of the clustering in redshift space. A quantitative assessment can be found in~\citet{Crocce:13}, where the multipoles of the redshift-space power spectrum are presented.
Figures~\ref{fig:RSD_multi_19_0} and~\ref{fig:no_RSD_multi_19_0} show the correlation function $\xi(r_{p},r_{\pi})$ of the total (left panels), central (centre panels) and satellite (right panels) galaxies brighter than $M_{r}< -19.0$  in redshift (top) and real (bottom) space, respectively. The plots are colour coded according to the amplitude values of the two-point correlation function. 
In Figure~\ref{fig:no_RSD_multi_19_0} one can observe that at the same luminosity threshold ($M_{r}< -19.0$) satellite galaxies (right panel) are more clustered than central galaxies. The whole population (centrals + satellites) has a clustering amplitude in between centrals and satellites, but closer to that of centrals as there are more central galaxies than satellites galaxies in this luminosity threshold sample (see Figure~\ref{fig:fraction_objects}).

In real space, Figure~\ref{fig:no_RSD_multi_19_0}, the clustering along the line of sight, $r_{\pi}$ and perpendicular to the line of sight, ($r_{p}$) are indistinguishable when taking into account the errors in the correlations. In redshift space, Figure~\ref{fig:RSD_multi_19_0}, the effects of galaxy peculiar velocities make the transversal and line of sight correlations  different. In the left and right panels one can observe both the Fingers-of-God and Kaiser effects. However in the central panel, where only the correlation function of central galaxies is shown, the Fingers-of-God effect is not present as we place centrals at the centre of their haloes without peculiar velocities. So, central galaxies only experience the large scale Kaiser effect.

It is also interesting to see the effect of galaxy bias in the redshift space clustering. In the context of linear theory and assuming the distant observer approximation, one can find an expression that relates the Fourier transform of the redshift space overdensity, $\tilde{\delta}_{s}(\vec{k})$, and the real space overdensity, $\tilde{\delta}(\vec{k})$ \citep[e.g.,][]{Dodelson:03}:
\begin{equation}
\tilde{\delta}_{s}(\vec{k})=\left[1+f\mu_{\boldsymbol{k}}\right]\tilde{\delta}(\vec{k})
\end{equation}
where $\mu_{\boldsymbol{k}}$ is defined to be $\hat{z}\cdot\hat{k}$, the cosine of the angle between the line of sight and the wavevector $\hat{k}$. Then, one can relate the power spectrum in redshift space, $P_{s}(\vec{k})$, and real space, $P(\vec{k})$:

\begin{equation}\label{eq:RSD_Pk}
P_{s}(\vec{k})=P(\vec{k})\left[1+\beta\mu_{k}^2\right]^2
\end{equation}
where it follows that $P_{s}(\vec{k})$ depends not only on the magnitude of $\vec{k}$ but also on its direction. The parameter $\beta$ is the linear growth rate, $f=\frac{d\ln{D}}{d\ln{a}}$ divided by the sample bias, $b_g$,  defined as $\delta_{g}\equiv b_{g}\delta$.  The correction due to redshift space distortions in equation~\ref{eq:RSD_Pk} is thus monotonic with $\beta$. Therefore, the smaller the galaxy bias, the stronger the correction to the power spectrum and equivalently the two-point correlation function. Satellite galaxies are always more clustered than central galaxies at a given luminosity and therefore central galaxies should experience a stronger Kaiser effect, as observed in Figure~\ref{fig:RSD_multi_19_0}.

\subsection{Galaxy colours}
\label{sec:assign_colour}

So far, we have dealt with the position and luminosities of galaxies, making sure that the clustering as a function of luminosity in our mock catalogue is as observed. Luminosity is a proxy to first approximation of the total number of stars in a galaxy. We now move to a property that is indicative of the type of stars that form a galaxy. The galaxy colour reflects the slope of the spectral energy distribution of the galaxy.  
This section describes the way we assign colours to galaxies in order to fit the observed $(g-r)$ .vs. $M_{r}$ distribution in the SDSS data~\citep[e.g.,][]{Blanton:03b} and the clustering properties as a function of colour Z11. \citet{Skibba:09} presented a method to incorporate colours in the HOD model in order to achieve these same goals.
Given the observed bimodality of the galaxy distribution in the colour-magnitude diagram~\citep[CMD,][]{Strateva:01,Baldry:04}, they start by parametrizing the colour distribution of red and blue galaxies as a function of luminosity. Then their procedure draws colours for galaxies from these distributions with probabilities depending on their luminosity and whether they are centrals or satellites.
We basically follow their method with some modifications that we describe below. 

\begin{figure}
	\centering
	\includegraphics[width=8cm]{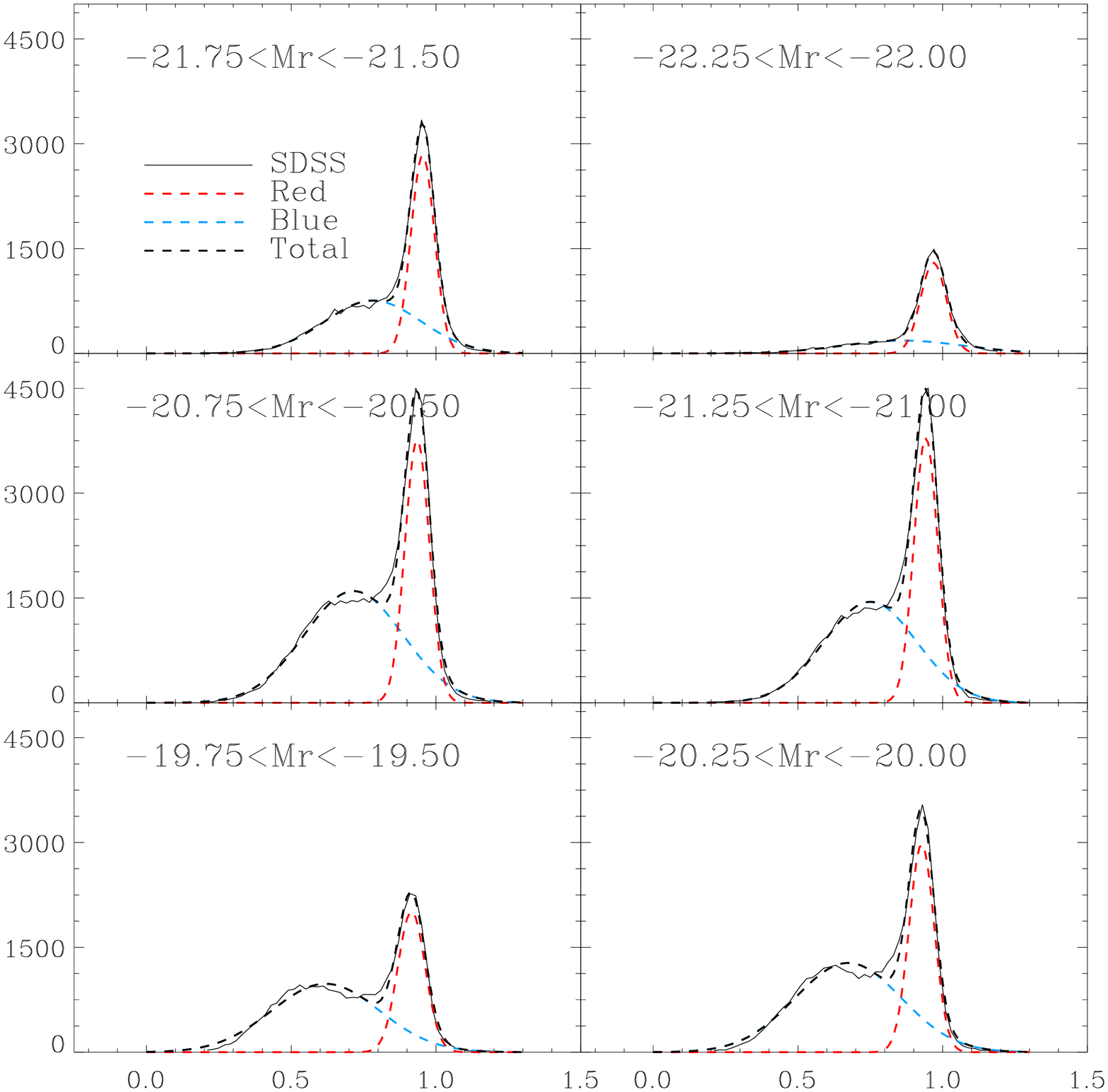}
	\caption{\small{Bimodal distribution of $(g - r)$ colour in the SDSS. The black solid line is the SDSS distribution and black dashed line is the sum of two Gaussian functions to the red and blue sequences (dashed lines).}}
	\label{fig:bimodal_SDSS_2_gauss}
	\includegraphics[width=8cm]{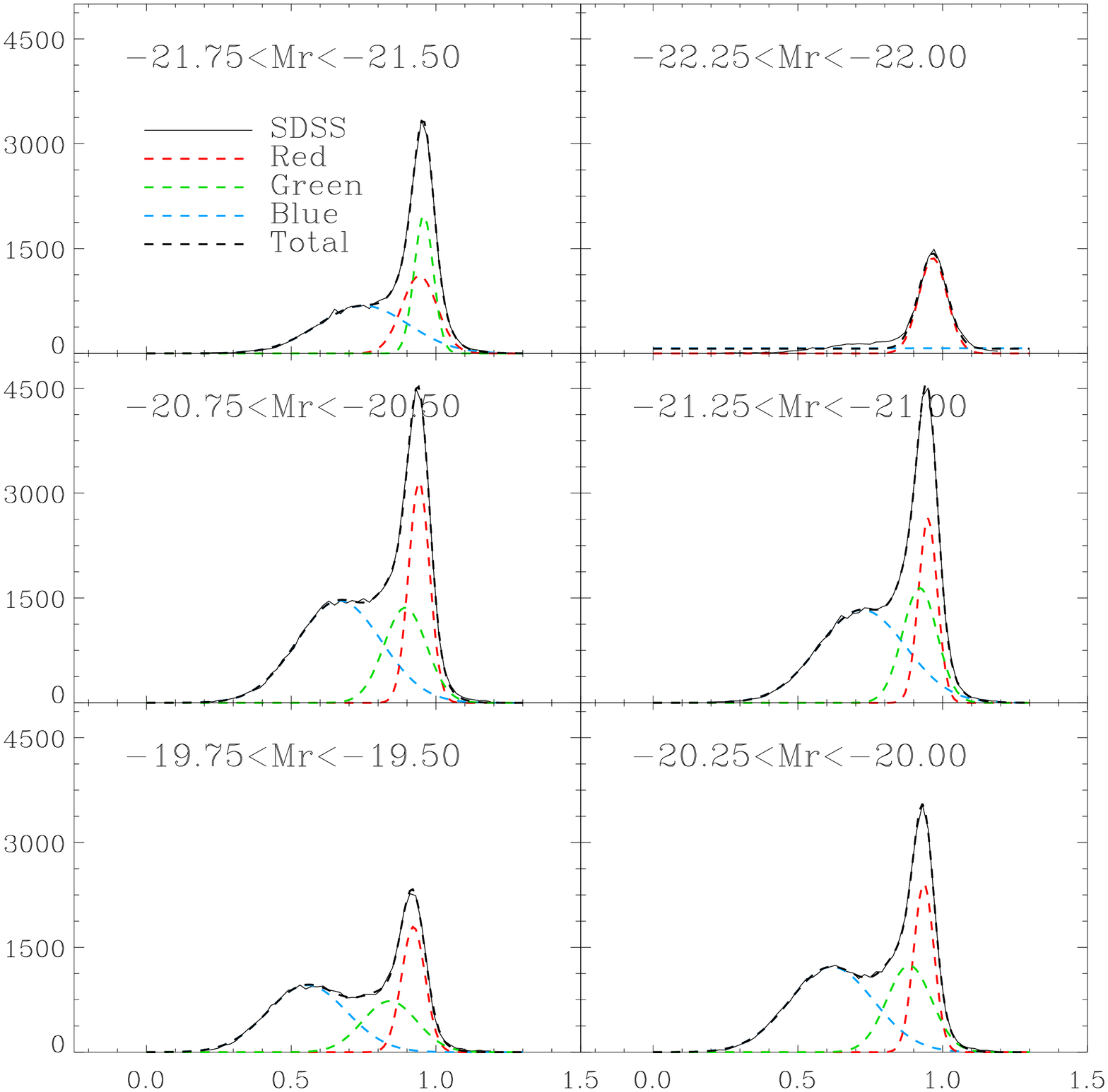}
	\caption{\small{SDSS (g-r) colour distribution with fits to three Gaussians as a function of luminosity, instead of the two Gaussians used in Figure~\ref{fig:bimodal_SDSS_2_gauss}.}}
	\label{fig:bimodal_SDSS_3_gauss}
\end{figure}

We start by pamaretrizing the colour distribution as a function of luminosity in the observed CMD
with the NYU DR7 catalogue~\citep{Blanton:05c}.
Figure~\ref{fig:bimodal_SDSS_2_gauss} shows the bimodality in the colour-magnitude diagram, $(g-r)$ .vs. $M_{r}$, by using different histograms of the number of galaxies as a function of colour for different luminosity bins. Following the model proposed by \citet{Skibba:09}, we fit two Gaussian functions for both, the blue and the red sequences. The solid black line represents the SDSS colour distribution for all galaxies, while the black dashed line corresponds to the sum of the two fitted Gaussian components, which correspond to the red and blue galaxy populations (red and blue dashed lines respectively). Comparing observations (solid black line) to the two Gaussians fit, we find that there is a mismatch in the number of galaxies between the two distributions. There is a small deficit of galaxies in the colour region in between the peaks of the red and blue distributions (the so called \textit{green valley}) when fitting only two Gaussian functions compared to observations. 

In order to better fit observations, we have decided to add another population in the fit. This will also give us somewhat more freedom when trying to adjust the clustering as a function of colour. We thus have red, green and blue galaxy populations. We proceed as before producing colour histograms in luminosity bins that we fit to three Gaussian distributions. We then fit the values of the mean and standard deviations of those distributions as a function of luminosity. The resulting relations are equations~\ref{eq:my_red_color_3gauss}, \ref{eq:my_green_color_3gauss} and \ref{eq:my_blue_color_3gauss}. 

\begin{equation}\label{eq:my_red_color_3gauss}
	\begin{array}{ll}
		\left< g-r | M_{r}\right>_{red} = 0.923-0.021(M_{r}+20.0) \\
		\sigma\left(g-r | M_{r}\right)_{red} = 0.041-0.003(M_{r}+20.0)
	\end{array}
\end{equation}

\begin{equation}\label{eq:my_green_color_3gauss}
	\begin{array}{ll}
                   \left< g-r | M_{r}\right>_{green} = \\
                   = 0.880-0.035(M_{r}+20.0)-0.062\tanh{\left(\frac{M_{r}+22.60}{0.12}\right)} \\
                   \sigma\left(g-r | M_{r}\right)_{green}= \\
                   = 
		  \begin{cases}
		0.055+0.023(M_{r}+20.0) \ \ \textrm{ if } M_{r} \geqslant -22.5\\ 
		-0.002-0.280(M_{r}+22.5) \textrm{ if } -22.5 >M_{r} \geqslant -22.75\\
		0.068 -0.020(M_{r}+22.75) \ \ \textrm{otherwise} \\
		\end{cases}
	\end{array}
\end{equation}

\begin{equation}\label{eq:my_blue_color_3gauss}
	\begin{array}{ll}
		\left< g-r | M_{r}\right>_{blue} =
		\begin{cases}
		0.610-0.100(M_{r}+20.0) \textrm{ if } M_{r} \geqslant -22.0\\ 
		0.810-0.020(M_{r}+22.0) \ \textrm{otherwise} \\
		\end{cases} \\
		\sigma\left(g-r | M_{r}\right)_{blue} = \\ = 0.170-0.005(M_{r}+20.0)-0.037\tanh{\left(\frac{M_{r}+22.38}{0.27}\right)}
	\end{array}
\end{equation}

\begin{figure}
	\centering
	\includegraphics[angle=0, width=8cm]{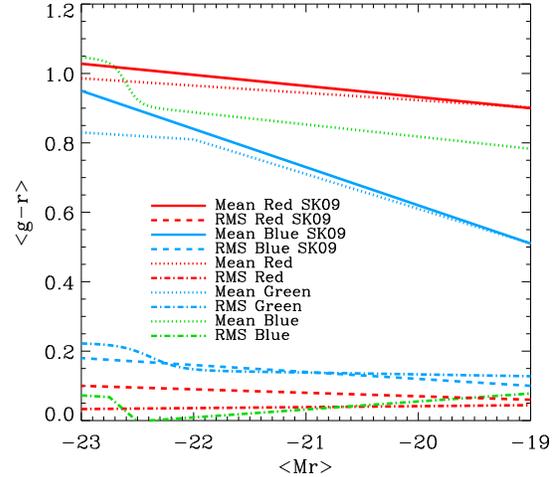}
	\caption{\small{Parameters found when fitting three Gaussian functions compared to the fits of \citet{Skibba:09} who used only two Gaussian functions. The red and blue solid and dashed lines are their fits to the mean and rms values respectively. The red, blue and green point-lines and point-dashed lines are the fits to our best mean and rms values when fitting three Gaussian functions.}}
	\label{fig:best_param_3gauss}
\end{figure}

\begin{figure*}
	\centering
	\includegraphics[width=16cm]{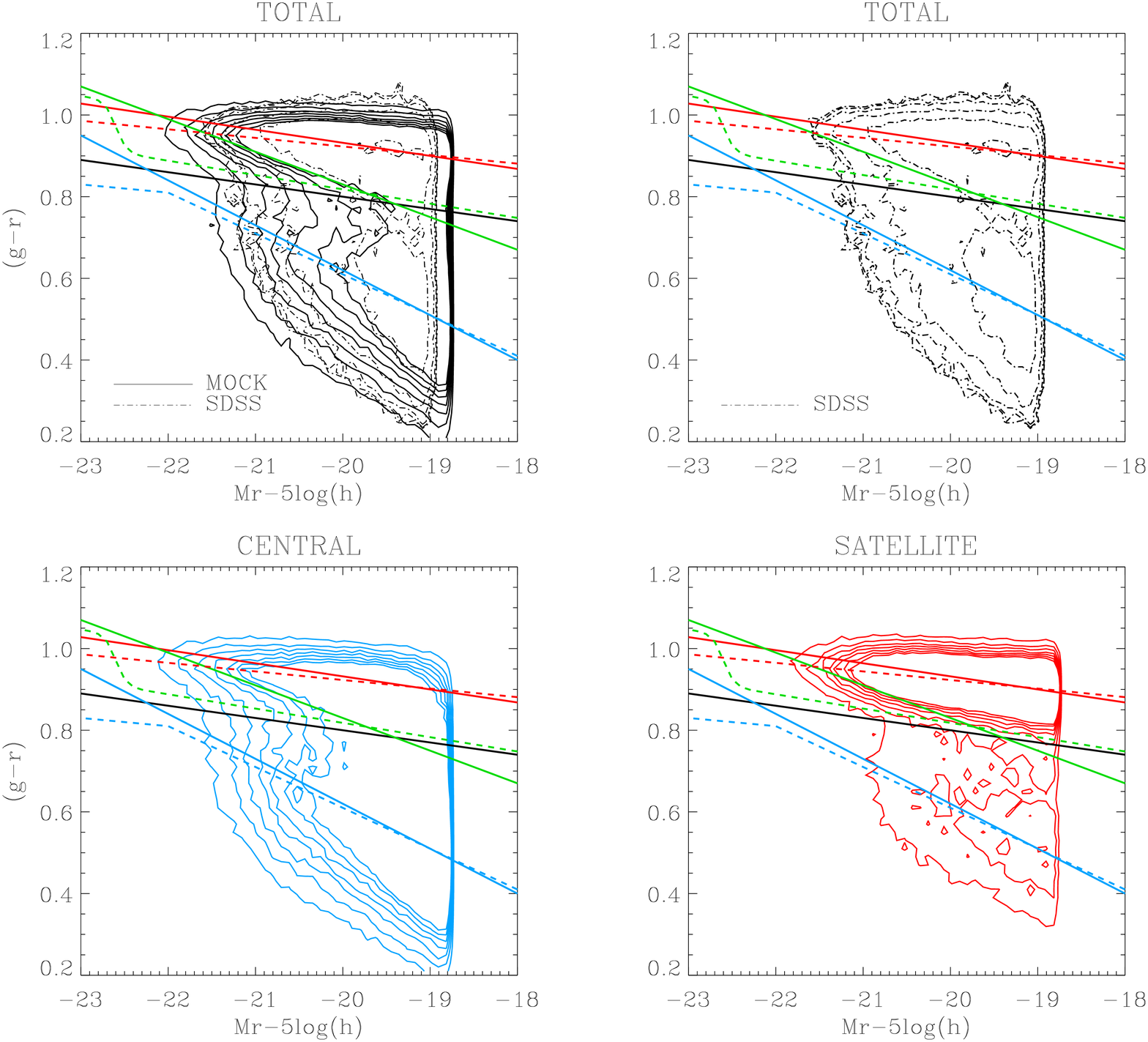}
	\caption{\small{Colour-magnitude diagram. The colour-magnitude diagram in the $M_{r}<-19.0$ SDSS volume limited sample is shown in the top right panel. The colour-magnitude diagram of the whole mock catalogue (central + satellite galaxies) is shown (solid black contours). The same contours as in the top right panel is show for comparison. The red and blue solid lines show the mean values of the red and blue sequences derived by \citet{Skibba:09}. The black solid line is a crude cut which some authors use to separate the galaxy population into the red and blue sequences. The green solid line refers to the satellite sequence set by \citet{Skibba:09}. The red, green and blue dashed lines are the mean values of the red, green and blue sequences we use to improve the fitting.}}
	\label{fig:CM_diagram}
\end{figure*}

Analogous to Figure~\ref{fig:bimodal_SDSS_2_gauss}, in Figure~\ref{fig:bimodal_SDSS_3_gauss} we show the SDSS colour distribution compared to our fit with three Gaussians functions. As expected the agreement is now much better. Figure~\ref{fig:best_param_3gauss} shows the luminosity dependence of the three colour sequences given by equations \ref{eq:my_red_color_3gauss}, \ref{eq:my_green_color_3gauss} and \ref{eq:my_blue_color_3gauss}.
The red sequence is rather similar in the range of luminosities probed. Its mean colour and standard deviations values change slightly with absolute magnitude, the distribution being somewhat wider and bluer at fainter luminosities.
The blue sequence has a much stronger mean colour dependence on luminosity and a wider distribution. 
The standard deviation is fairly constant across luminosity. The behaviour is different at bright absolute magnitudes. The number of blue galaxies is scarce here (Figure~\ref{fig:CM_diagram}) and the absence of a green population makes the standard deviation increase. The green population is peculiar in the sense that we have joint two different populations in one. On the one hand, there is the proper green population at absolute magnitudes fainter than $M_{r}>-22.5$, whose mean colour is in between the red and blue sequences and gets redder with luminosity. Its standard deviation diminishes with luminosity up to the point $M_{r}\sim-22.3$ where there are no more galaxies belonging to the green sequence. 
On the other hand, there is a population of redder objects than the main red sequence at bright luminosities with a wider red colour tail. This population is included in the green sequence fit of equations~\ref{eq:my_green_color_3gauss} for convenience, although it is clearly distinct than the proper green sequence. We speculate that this population can correspond to galaxies with higher metallicity or to dustier galaxies or simply be due to larger errors in the photometry. 

\begin{figure}
	\centering
	\includegraphics[angle=0, width=8cm]{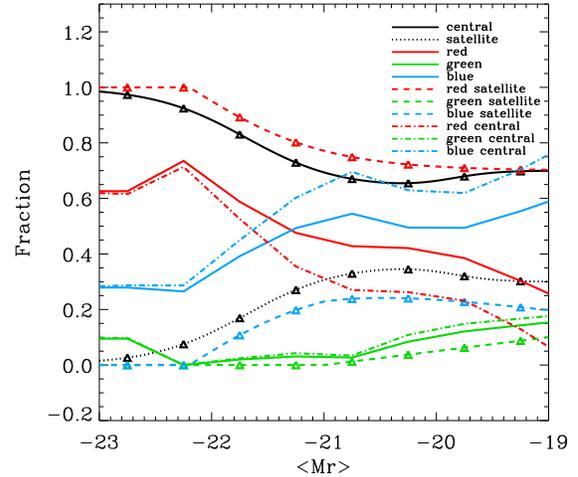}
	\caption{\small{Fraction of objects as a function of the absolute magnitude $M_{r}$. The black solid line refers to the fraction of central objects. The black dotted line is the satellite fraction of objects. The red, green and blue solid lines correspond to the total (central + satellite) red, green and blue fraction of galaxies. The dashed and dashed dotted lines show the same colour fractions but for central and satellite galaxies respectively.}}
	\label{fig:fraction_objects}
\end{figure}

\begin{figure*}
	\centering
	\includegraphics[width=16cm]{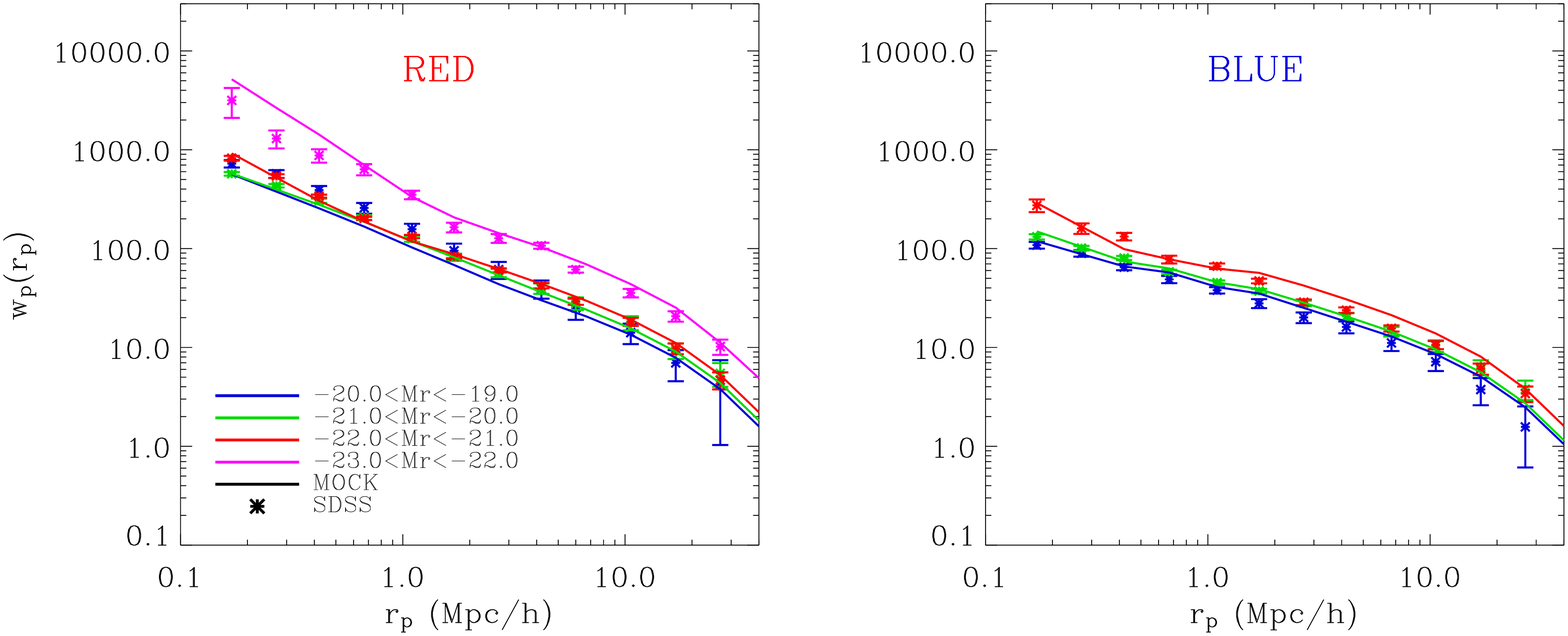}
	\caption{\small{Projected correlation function for different luminosity-bin samples of the catalogue. The left panel shows red galaxies and the right panel shows blue galaxies. The brightest blue sample of galaxies is not present since their correlation function is noisy.}}
	\label{fig:wp_color_bins_plot}
\end{figure*}

The way we assign colours to galaxies is similar to the model presented in \citet{Skibba:09} in the sense that it depends on the galaxy type (whether it is a central or a satellite galaxy) and also on the colour sequence (blue, green or red) it belongs to. We also assume that colours do not depend directly on the halo mass. First we assign colours to satellite galaxies and later we force central colours to follow the total colour-magnitude diagram. \citet{Skibba:09} proposed an ad-hoc function that defines the mean colour of satellite galaxies given their luminosity. They used this function to define the fraction of satellite galaxies that belongs to the red sequence. In our case, we directly set the fraction of satellite galaxies that belongs to the red and green sequences as a function of absolute magnitude (equations~\ref{eq:f_sat_red} and~\ref{eq:f_sat_green} and red and green dashed lines in Figure~\ref{fig:fraction_objects} respectively). Once these function are set, the fraction of blue satellites is fixed  (equation~\ref{eq:f_sat_blue}). Taking into account the HOD, then the fractions of red, green and blue centrals follow to yield equations~\ref{eq:fcenred}, \ref{eq:fcengreen} and~\ref{eq:fcenblue}. We modify the red and green satellite fraction  functions until we find an acceptable fit to the galaxy clustering as a function of colour and luminosity. The freedom that two arbitrary functions provide is large and difficult to sample. We have thus started from the red satellite fraction function of~\citet{Skibba:09} and modify it slightly until convergence to the observed clustering is achieved. The overall contribution of the green sequence to the clustering is small given its small relative number compared to red and blue galaxies. This overall small green galaxy fraction imposes that the green satellite fraction needs to be also small. We have thus only tried small values for the green satellite fraction with mild luminosity dependence slopes in our comparisons to observations. The modifications of these red and green satellite fractions mostly affect the one-halo term regime of the clustering. The two-halo term regime is minimally altered as we have assumed that the colour assignation does not depend on halo mass which is the property that would allow us to change the large scale clustering.
In more detail, we separate the red and green fractions functions into luminosity bins. We change the values of the red and green fraction in each bin, generate a catalogue with that recipe and compute the correlation function as a function of colour and luminosity of the resulting catalogue. We compute a diagonal-$\chi^2$ of the clustering of the catalogue compared to the observations of Z11\footnote{As before, we have not included covariance between different scales and therefore it is not a proper $\chi^2$. We change the values of the fractions as a function of the luminosity until we find acceptable fits. For convenience, we also impose that the fractions as a function of luminosity change monotonically with luminosity as to be able to fit an analytic function. Note that although it may seem that there can be a lot of freedom in choosing these fractions, in fact the values that they can take are constrained by the colour-magnitude diagram and the HOD. We run our minimization until we reached values of our diagonal-$\chi^2$ per degree of freedom of 1 and 2, respectively for the red and blue galaxies clustering as a function of luminosity. We find difficult to improve the fit of the clustering of the blue galaxies without introducing further dependences on the galaxy fractions. This is due to the clustering of the blue galaxies luminosity bin $-22<M_r<-21$ which is the only one that does not fit properly (see Figure~\ref{fig:wp_color_bins_plot}).} 
Overall, we find the following expressions\footnote{In these equations for the fractions of galaxies whenever a value is larger than 1 or lower than 0, it is set to 1 and 0 respectively.}.

\begin{equation}\label{eq:f_sat_red}
f_{sat}^{red}(M_{r})=1.00-0.30\tanh{\left( \frac{M_{r}+22.20}{1.20}\right)}
\end{equation}

\begin{equation}\label{eq:f_sat_green}
f_{sat}^{green}(M_{r})=0.05-0.05\left( M_{r} +20.0\right)
\end{equation}
\begin{equation}\label{eq:f_sat_blue}
f_{sat}^{blue}(M_{r})=1-\left( f_{sat}^{red}(M_{r}) + f_{sat}^{green}(M_{r})\right)
\end{equation}

The red, green and blue central galaxy fractions are fixed once we take into account the overall fraction of central and satellites as a function of luminosity given by the HOD prescription

\begin{equation}
f_{tot}^{red}(M_{r})=f_{cen}^{red}(M_{r})f_{cen}(M_{r}) + f_{sat}^{red}(M_{r})f_{sat}(M_{r})
\end{equation}
finding the following relations
\begin{equation}\label{eq:fcenred}
f_{cen}^{red}(M_{r})=\frac{f_{tot}^{red}(M_{r}) + f_{sat}^{red}(M_{r})f_{sat}(M_{r})}{f_{cen}(M_{r})}
\end{equation}

\begin{equation}\label{eq:fcengreen}
f_{cen}^{green}(M_{r})=\frac{f_{tot}^{green}(M_{r}) + f_{sat}^{green}(M_{r})f_{sat}(M_{r})}{f_{cen}(M_{r})}
\end{equation}

\begin{equation}\label{eq:fcenblue}
f_{cen}^{blue}(M_{r})=1-(f_{cen}^{red}(M_{r})+f_{cen}^{green}(M_{r}))
\end{equation}

In order to better visualize the relative contributions, in Figure~\ref{fig:fraction_objects} we show the fraction of objects of the different populations as a function of the absolute magnitude $M_{r}$. The black solid line and the black dotted line show the overall fraction of central and satellite galaxies respectively. 
The red, green and blue solid lines represent the fraction of red, green and blue galaxies for all galaxies.
The figure also shows the fraction of central (dot-dashed lines) and satellite galaxies (dashed lines)  that belong to the red, green and blue sequences. It is worth noticing that satellite galaxies are mostly red, reflecting the influence that environment has on galaxy evolution. Dense environments preclude star formation and thus galaxies tend to have redder colours. Another subtlety comes from the ``fourth'' luminous red population that we have ascribed to the green sequence. At these bright luminosities most of the galaxies are centrals. We have thus made this population be central galaxies. This makes the red and blue fractions trends look discontinuous at bright absolute magnitudes.  

In detail, our algorithm assigns colours to galaxies in the following way: for each satellite galaxy we draw two random numbers, $u_{0}$ and $u_{1}$.  $u_{0}$ is a uniformly distributed random number between 0 and 1, and determine to which sequence the satellite galaxy belongs to depending on the expressions \ref{eq:f_sat_red}, \ref{eq:f_sat_green} and \ref{eq:f_sat_blue}. Then, we compute the mean value of the colour depending on the luminosity following the expressions \ref{eq:my_red_color_3gauss}, \ref{eq:my_green_color_3gauss} or \ref{eq:my_blue_color_3gauss}. $u_{1}$ follows a Gaussian distribution with mean value 0 and rms = 1 and is used to generate a realization of the colour.

\begin{equation}
(g-r)_{realization} = (g-r)_{mean} + u_{1} \;\sigma(g-r)
\end{equation}
For central galaxies, we proceed in the same way but using expressions \ref{eq:fcenred}, \ref{eq:fcengreen} and \ref{eq:fcenblue} to determine to which sequence the central galaxy belongs to. The first random number determines the galaxy sequence  and the second random number is used to generate a realization of the colour given the sequence it belongs to.

The colour-magnitude diagram in the $M_{r}<-19.0$ SDSS volume limited sample (black dashed-dot contours) is shown in the top left panel of Figure~\ref{fig:CM_diagram}, while the top right panel shows the colour-magnitude diagram for all (central + satellite) galaxies of the mock galaxy catalogue (for better comparison, we also show it in the  top left panel). The red and blue solid lines are the mean of the colour distributions of red and blue galaxies presented in \citet{Skibba:09} in their fits to the SDSS data using two Gaussian functions. In dashed lines we present our fits to the mean of the red, green and blue sequences when using three Gaussian functions in order to model the colour distribution as a function of absolute magnitude. The green solid line refers to the satellite sequence set by \citet{Skibba:09} (their equation~7). By construction, the agreement between the SDSS DR7 data and the galaxy mock is very good. The colour-magnitude diagram for the mock galaxy catalogue for central (bottom left panel) and satellite (bottom right panel) galaxies is shown in the bottom left and right panels, respectively.

Figure~\ref{fig:wp_color_bins_plot} shows the colour dependence of the clustering of the catalogue with the projected two point correlation function. We show separately red (left panel) and blue galaxies (right panel). Here, red and blue populations are separated using the criterion of Z11 and not our separation to assign colours. Data points with error bars correspond to SDSS data from Z11 and the lines refer to the mock catalogue. For easier comparison,  we have split the catalogue into the same four luminosity threshold samples of Z11. We have also adopted the same spatial binning. The observed correlation function of the brightest blue galaxies is not presented because it is very noisy given the very few objects. Red galaxies are more clustered than blue galaxies at all scales. However, at the one-halo term scales, the clustering difference is much stronger. Satellite galaxies dictate the clustering at small scales and as we have seen (Figure~\ref{fig:fraction_objects}), most satellites are red.
The three dimensional correlation function of central galaxies is zero at the one-halo term scales due to the halo exclusion effect. However, in the projected correlation function the correlation strength is flat at those scales (Figure~\ref{fig:wp_th_bins_plot_cen_sat}, left panel). The blue satellites are a minor fraction of the total blue galaxies, typically around 15\% (although it is a function of luminosity), and therefore their contribution to the clustering is small, resulting in the projected correlation function staying flatter than red galaxies at those scales. 
At the two-halo scales, the correlation of our blue galaxies is somewhat more strongly clustered than observations. We have assumed that colours only depend on absolute magnitude. For centrals, this is the same as a dependence on halo mass except at the luminosities where scatter in the halo mass-luminosity is important. However, for satellites this is not the case. Our results seem to indicate that satellite colours should have some dependence on halo mass. If blue satellites populate haloes of lower mass on average then their clustering amplitude would be reduced. 
This tendency is consistent with halo assembly bias~\citep[e.g.,][]{Gao:05,Wechsler:06,Gao:07,Li:08}, although we phrase it in terms of halo mass instead of halo assembly time as we do not retain that information in our simulation lightcone output.

\begin{figure*}
	\centering
	\includegraphics[width=14cm]{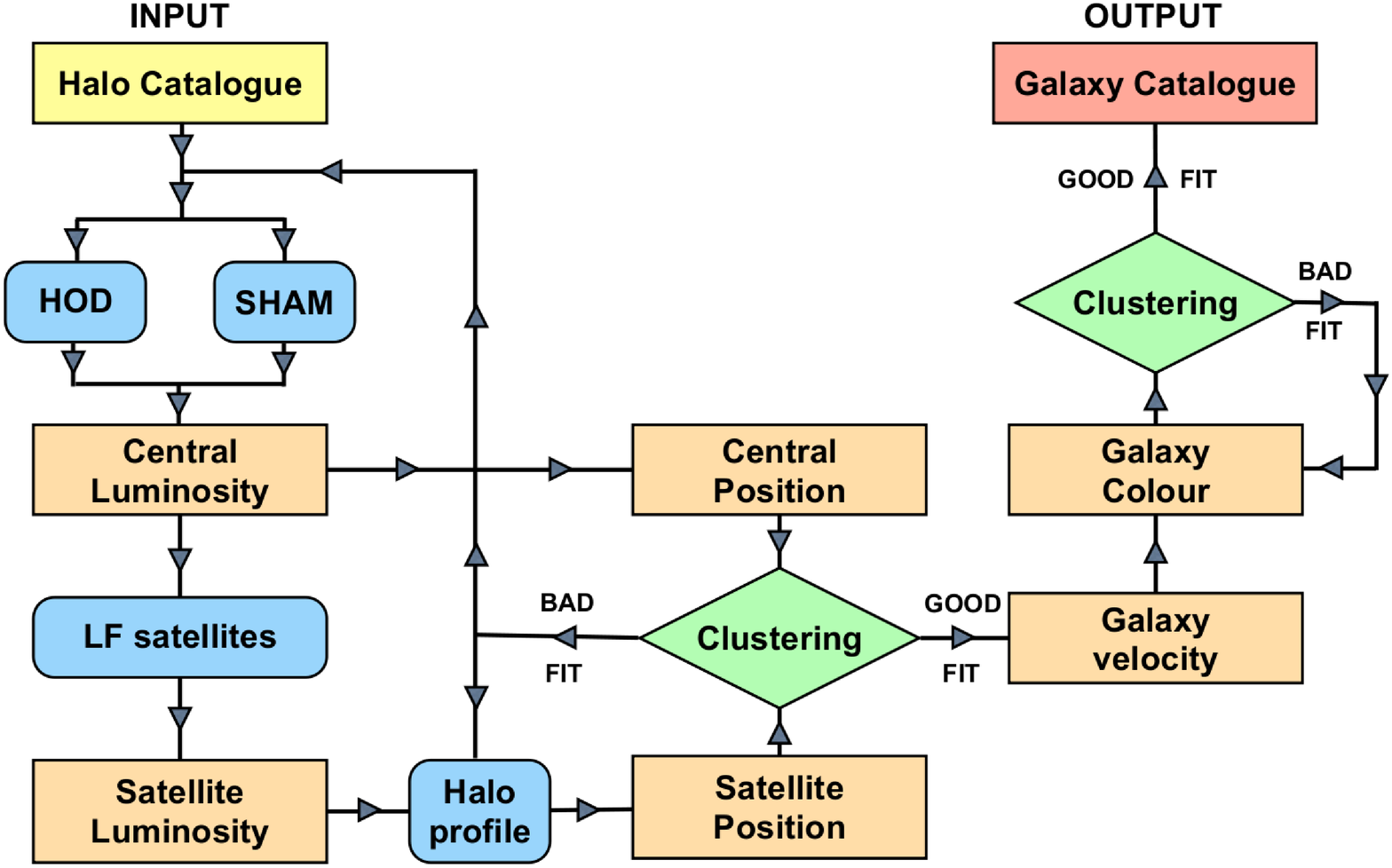}
	\caption{\small{Simplified flow diagram of the mock catalogue construction algorithm, colour coded depending on the functionality of each module. In yellow, we have the input catalogue. In blue, modules that require modelling. In orange, we assign observed properties to the galaxies. In green, we check the clustering properties. And in red the output catalogue. Note that this is a simplified version and for a proper description you should refer to the text.}}
	%\caption{\small{Schema of the algorithm. \textcolor{blue}{It is necessary to explain it.}}}
	\label{fig:algorithm}
\end{figure*}

In the right panel of Figure~\ref{fig:gal_mean_bias_Mr_CF_3D_total_colour} we present the large scale bias as a function of $M_{r}$ for red (triangles) and blue (asterisks) galaxies in the same way as we previously shown in section \ref{sec:fine_tuning} for total, central and satellite galaxies. As seen before, red galaxies are more clustered than blue ones at all luminosities as expected. Blue galaxies at low luminosities are anti-biased as the less massive haloes (left panel of Figure~\ref{fig:gal_mean_bias_Mr_CF_3D_cen_sat}). This is an indication that most faint blue galaxies are central galaxies in low mass haloes as expected from Figures~\ref{Mh-Lcen_rel} and~\ref{fig:fraction_objects}.

\section{Conclusions}
\label{sec:conclusions}

In recent years, there has been a lot of effort to design and carry out large galaxy surveys to constrain and understand our cosmological model and in particular the nature of dark energy. Large numerical simulations are needed to interpret the results from these surveys. The MICE project is producing simulations to help in this scientific enterprise. In particular, the MICE Grand Challenge is one of the largest N-body simulations run to date that produces dark matter, halo and galaxy catalogues including lensing properties. The outputs of the simulation are public through a dedicated web portal (http://cosmohub.pic.es) and are described in a series of papers~\citep{Fosalba:13a, Crocce:13,Fosalba:13b, Hoffmann:14}. We now complement those papers presenting the method we have used to construct the galaxy catalogues. Here, we present our basic methodology applied to a local snapshot output where the observational constraints are tight to optimize our recipes. In future work (Castander et al., in prep), we will present how we extend our catalogues to the lightcone to produce a realization of the observed universe. In this paper, we emphasize the methodology itself trying to understand why we need to introduce our recipes and what their effect is on the clustering properties of the catalogue. Given the many degrees of freedom, it is unfeasible to produce a proper optimization sampling of all the parameter space introduced. We explore the possible values of our parameters and functions until a reasonable fit is found without requiring statistical minimization. We present the values used in the catalogue released in our web portal. In some cases, we have later found better parametrizations or functions that improve the fit to observations. We have opted to present the values of the release, though.

We present a method to generate a mock galaxy catalogue starting from a halo catalogue. We optimize the method populating the redshift zero snapshot halo catalogue of the MICE-GC simulation and comparing it to observational constraints from the SDSS. We modify the parameters and functions of our method to reproduce the observed galaxy luminosity function, the colour-magnitude diagram and the clustering as a function of luminosity and colour. The main properties that we simulate are then positions, velocities, luminosities and colours. 
Figure~\ref{fig:algorithm} provides a glimpse of our algorithm that we summarize in the following points:

\begin{itemize}

\item{Following the Halo Occupation Distribution philosophy, we assume that galaxies populate dark matter haloes and its population can be divided into central and satellite galaxies.}

\item{We use a halo catalogue to produce galaxies using only information on the mass, position and velocity of each halo. We ignore the substructure inside the dark matter haloes and in particular subhaloes. We use the HOD parametrization to generate satellites and in this way we can push fainter into the galaxy LF using few particles friend-of-friends associations. We can therefore generate highly complete mock galaxy catalogue for deep cosmological surveys.}

\item{We assume that each halo contains one central galaxy located at its centre and a number of satellites given by a Poissonian realization of the mean occupation number given by the HOD parametrization.}

\item{We generate a cumulative galaxy number function as a function of halo mass using the HOD prescription. Following the abundance matching concept, we compare this function to the cumulative galaxy luminosity function to obtain a relation between halo mass and luminosity. We use this relation to assign luminosities to the central galaxies given the halo mass they populate.}

\item{We compute a cumulative luminosity function for satellite galaxies subtracting the cumulative LF for centrals from the overall galaxy luminosity function. We assign luminosities to satellites drawing them at random from this cumulative LF.}

\item{We introduce scatter in the halo mass-luminosity relation to reduce the clustering amplitude in the two-halo term scales at bright luminosities.}

\item{We distribute satellites following an NFW profile. However, in order to reproduce the observed one-halo term clustering, we need to concentrate galaxies more than expected using standard relations between the halo mass and the concentration parameter.}

\item{In order to better reproduce simulations results and observational data, we use triaxial halo shapes to place our galaxies.}

\item{We assume that central galaxies are at rest at the halo centre and assign them the mean halo velocity. We assign peculiar velocities to satellite galaxies assuming that they are Gaussian distributed with a velocity dispersion given by the halo mass.}

\item{We assign colours to our galaxies using a similar procedure to that of~\citet{Skibba:09}. We parametrize the colour-magnitude diagram with three Gaussian populations\footnote{In fact, we distinguish four populations but we incorporate the red luminous subpopulation into the green population.} in colour (red, green and blue galaxies) whose mean and standard deviations vary as a function of luminosity. We then draw colours for galaxies from these distributions depending on their luminosities and whether they are centrals or satellites. For that purpose, we generate two functions giving the probability that a satellite belongs to the red and green sequences. From these functions, the rest of the needed probability functions to assign colours are derived.}

\end{itemize}

The main goal of the work described in this paper is to build galaxy catalogues that can be used to design and interpret cosmological surveys. We use HOD and SHAM techniques to populate dark mater haloes with galaxies. We assign positions, velocities, luminosities and colours to the galaxies in a way that the final catalogue reproduces the main observables. In the process, we have made some assumptions and applied some prescriptions in order to achieve a good fit to observations. The main lessons learnt in the process can be summarized as follows. 

Normally, the HOD model is used to interpret the clustering properties of luminosity threshold samples inferring the halo masses that galaxies populate. Different samples require different values of the HOD parameters that give the halo occupation. In our catalogue building procedure, we need to address the inverse problem. Knowing the halo masses, we want to infer the galaxy luminosities. As in the standard HOD usage, we find that a single parametrization cannot fit the  clustering for all luminosities. We have thus chosen to fit a function to one of the HOD parameters, providing the flexibility needed to fit the data.

After~\citet{Zheng:05}, most HOD parametrization include a fuzzy cut in the occupation number of centrals, in the sense that for a luminosity threshold sample the sharp cut in luminosity does not correspond to a sharp cut in mass of the haloes containing galaxies. In the case of ~\citet{Zheng:05}, they use an error ($\erf$) function (their equation~1) to give a varying probability for a halo to contain a central galaxy as a function of the lowest halo masses. We find that when the luminosity limit corresponds to halo masses below $\log (M_{h}) \sim 12.0$, the bias of the lowest mass haloes is approximately constant with halo mass and therefore the fuzzy limit is ineffective to change the clustering properties of the sample and thus not needed for our purposes.

The HOD paradigm establishes that halo mass is the property that dictates the clustering of the galaxy population. Placing satellites inside the haloes, we find that we need a dependence on another parameter to better fit the one-halo term clustering. This is generally referred to as halo assembly bias and usually discussed in the literature in term of the satellite galaxies assembly times. Here, we discuss the effect in terms of their luminosities as we do not retain the assembly information in our lightcone halo catalogue.

In order to be complete down to faint magnitudes, our method pushes the lowest halo mass that we can use to friend-of-friends associations down to ten particles. While it is arguable that these associations represent real haloes,  
higher mass resolution simulations indicate that the bias of haloes in this low mass range does not appreciably depend on mass. As we correct the masses of these few particles associations to coincide with the abundance of haloes in higher resolution simulations, we expect the resulting clustering to be adequate. In fact, when we populate these haloes with galaxies, we find that we can reproduce the observe clustering, validating the procedure. We defer a more detail investigation of this issue to further work.

Centrals are the dominant population in a luminosity threshold galaxy sample. In the luminosity range sampled by our catalogue, central galaxies comprise approximately 70\% of the galaxy population. At the brightest luminosity end, this fraction is even larger with the majority of the most luminous galaxies being centrals.

Central galaxies are the dominant contributors to the clustering amplitude in the two-halo term regime. At these scales and for the same luminosity range, satellites populate more massive haloes than centrals and therefore are always more clustered than centrals, but their overall contribution is lower given their fewer number (Figures~\ref{fig:wp_th_bins_plot_cen_sat} and~\ref{fig:fraction_objects}).

The distribution of satellite galaxies inside their haloes defines the clustering properties in the one-halo regime.
We find that we need to place satellites closer to the halo centre than naively expected. Algorithmically, we implement it as a change in the concentration parameter. Moreover, our fits to the observed clustering improves if the change in the concentration depends on luminosity. For central galaxies a dependence on luminosity is equivalent to a dependence on halo mass, as there is a one-to-one relation between the two, except the scatter at high halo masses. However, for satellites this is not the case, as the satellites in a given halo have a range of luminosities and also at a given luminosity there are satellites in haloes spanning a wide range of halo masses. The indication that the position of satellites has some dependence on their luminosity may be a sign of luminosity segregation in clusters of galaxies and would break the HOD assumption that the only relevant property is halo mass.

In our method, the assignation of luminosities to satellite galaxies is done drawing them randomly from the overall satellite cumulative luminosity function, with no dependence on the halo mass. In the released catalogue we include an extra constraint  with a penalization recipe to modify the two-halo term clustering. In general, adding an extra dependence on halo mass in this assignation provide us with an extra degree of freedom to improve the fits to observations. However, there are many parameters in the method and there are instances when extra parameters to improve the clustering are not necessary.
Equivalently, the assignation of colours to satellite galaxies only depends on their luminosity, and not on the halo mass they populate. We find that the two-halo term clustering of the blue galaxies is somewhat stronger than observed, suggesting that these blue satellite probably populate less massive clusters than the ones randomly assigned. This is a trend equivalent to the results of the age abundance matching of~\citet{Hearin:13b}.

We divide galaxies into three distinct colour populations characterized by Gaussians in the colour magnitude diagram. 
We find that three populations provide a better fit to the data than the typical two populations used.  We provide fits to the mean colour and standard deviations of these populations. Our ``green'' population fit is in fact a fit to two populations. The ``proper'' green population, with colours in between the red and blue populations and absolute magnitudes fainter than $M_r\sim-22.5$; and a red population, absorbed in our green population fit, at brighter magnitudes. The green galaxies are necessary to explain the excess of galaxies compared to the two Gaussian fit. For our purposes, it is indifferent whether this green sequence corresponds to a physically distinct type of galaxies or just to the inadequacy of the Gaussian distribution to represent the red and blue populations. At the most luminous absolute magnitudes the red galaxies become redder than the extrapolation of the mean of the red population and their standard deviation increases. This is an indication that the red population somehow changes its properties for the most luminous galaxies, probably due to the fact that these galaxies live in the densest environments that alter their evolution. We have arbitrarily opted to keep the trends with luminosity of the red galaxy fits and include these new red galaxies into the fit of the green population.

We assign the halo velocity to the central galaxies. For satellite galaxies, we assume that they can be represented by the halo velocity dispersion obtained from their mass. Here, we provide a qualitative description of the velocity assignation implications on clustering. \citet{Crocce:13} perform a more detailed study which suggests that our method produces results in accordance to theoretical expectations. While adequate for large scale clustering, our simple assignation
method may not be appropriate for cluster dynamics studies.

In this paper we have described a method to build mock galaxy catalogues from halo catalogues. We have applied it to the local snapshot of the MICE Grand Challenge simulation. We show that we are capable of reproducing the luminosity function and colour magnitude diagram observed by SDSS and the clustering dependence on luminosity and colour. In our next paper, we will present how we extend our method and apply it to the lightcone to generate a catalogue as will be observed by on going and future large cosmological galaxy surveys. This catalogue is available through the CosmoHUB web portal (http://cosmohub.pic.es).

\section*{Acknowledgements}

We would like to thank Anne Bauer, Carlton Baugh, Ravi Sheth, Ramin Skibba for useful discussions.
We also thank the referee, Martin White, for his constructive comments that improved the quality of the paper. JC acknowledges the hospitality of University of Pennsylvania where part of this work was developed.
FJC acknowledges the hospitality of Fermilab National Laboratory where part of this work was developed.
We are greatly indebted to Christian Neissner, Davide Piscia, Santi Serrano, Pau Tallada and Nadia Tonello for their development of the simulations webportal.
The MICE simulations have been developed at the MareNostrum supercomputer (BSC-CNS, www.bsc.es) thanks to grants AECT-2006-2-0011 through AECT-2010-1-0007. Data products have been stored at the Port d'Informaci\'o Cient\'{\i}fica (PIC, www.pic.es). Funding for this project was partially provided the European
Commission Marie Curie Initial Training Network CosmoComp
(PITN-GA-2009 238356),
the Spanish Ministerio de Ciencia e Innovacion (MICINN),
projects 200850I176, AYA-2006-06341, AYA-2009-13936, AYA-2012-39620, AYA-2012-39559, 
Consolider-Ingenio CSD2007-00060, research project SGR-1398
from Generalitat de Catalunya and the CSIC grant PA1002962. MC acknowledges support from the Ram{\'o}n
y Cajal MICINN program.

\bibliography{aamnem99,biblist}

\appendix
\section{Explicit halo mass function comparison}
\label{app:appendix1}

In section \ref{sec:clustering_constraints_centrals} we show the large scale MICE-GC halo bias and compare it with fits from \citet{Manera:10}. In this appendix we show the explicit relation between \citet{Crocce:10} and \citet{Manera:10} halo mass functions.

Halo bias can be measured in different ways (see e.g \citealt{Manera:10}). It measures the relation between the matter density field and the halo density field. The linear large scale halo bias, $b^{Lin}_{h}$, can be derived using the two point correlation function.

Expression \ref{halo_bias_cf} relates the two point correlation function of haloes and dark matter assuming a linear bias relation between the haloes and dark matter distributions, $\delta_{h}(\textbf{r}) = b_{h}^{Lin}\delta_{m}(\textbf{r})$:

\begin{equation}\label{halo_bias_cf}
\xi_{h}(r)=\left(b^{Lin}_{h}\right)^{2}\xi_{DM}(r)
\end{equation}

\citet{Manera:10} presented a maximum-likelihood method for fitting parametric functional forms to halo abundances. They showed a linear bias function for haloes applying the peak background split to a Warren-form halo mass function:

\begin{equation}\label{eq:Warren bias}
b^{W}\left(\nu\left(\sigma\left(M_{h}\right)\right)\right)=1+\frac{c^{'}\nu-1}{\delta_{c}}+\frac{2a^{'}b^{'}+b^{'}+\left(c^{'}\nu\right)^{a^{'}}}{\delta_{c}\left(b^{'}+\left(c^{'}\nu\right)^{a^{'}}\right)}
\end{equation}
where $\nu$ is the natural scaling variable according to the spherical evolution model and is defined as: 

\begin{equation}\label{Nu}
\nu\equiv\frac{\delta_{sc}^{2}\left(\Omega_{z},\Lambda_{z},z\right)}{\sigma^{2}(M_{h})}
\end{equation}
where $\delta_{sc}$ is the critical density required for spherical collapse given a cosmological model $\left(\Omega_{z},\Lambda_{z}\right)$:

\begin{equation}\label{delta_colapse}
\delta_{sc}\left(\Omega_{z},\Lambda_{z},z\right)=\frac{\delta_{sc}(0)}{D(z)}
\end{equation}
where $\delta_{sc}(z=0)=1.686$ in an Einstein-de Sitter cosmology, and $\delta_{sc}(z=0)=1.673$ in the MICE cosmology.

They also suggested an expression for the MF depending on the variable $\nu$:

\begin{multline}\label{eq:nu_fnu}
\nu f_{W}(\nu)=\frac{M_{h}}{\rho_{b}}\frac{dn(M_{h})}{d\ln{M_{h}}}\frac{d\ln{M_{h}}}{d\ln{\nu}}=\frac{M_{h}}{\rho_{b}}\frac{dn(M_{h})}{d\ln{\nu}}= \\ =A^{'}\left[1+b^{'}(c^{'}\nu)^{-a^{'}}\right]\exp{\left(-c^{'}\nu/2\right)}
\end{multline}

In order to derive the MICE parameters of halo bias we use the MF expressions from \citet{Crocce:10} and \citet{Manera:10}. We use the definition of the differential MF:

\begin{equation}\label{diffMF}
f(\sigma,z)=\frac{M_{h}}{\rho_{b}}\frac{dn\left(M_{h},z\right)}{d\ln{\sigma^{-1}\left(M_{h},z\right)}},
\end{equation}
equation~\ref{eq:nu_fnu} and the relation between $\nu$ and $\sigma$ (equation~\ref{Nu}) to get equation~\ref{fWarren_relation}:

\begin{equation}\label{fWarren_relation}
\nu f_{W}(\nu)=\frac{1}{2}f(\sigma,z)
\end{equation}

Parameters are related in this way:

\begin{equation}\label{c}
c=c^{'}\frac{\delta_{sc}^{2}}{2D^{2}} \qquad c^{'}=c\frac{2D^{2}}{\delta_{sc}^{2}}
\end{equation}
\begin{equation}\label{a}
a=-2a^{'} \qquad a^{'}=-\frac{a}{2}
\end{equation}
\begin{equation}\label{A}
A=2A^{'}b^{'}\left(c^{'}\frac{\delta_{sc}^{2}}{D^{2}}\right)^{-a^{'}} \qquad A^{'}=\frac{1}{2}Ab
\end{equation}
\begin{equation}\label{b}
b=\frac{1}{b^{'}}\left(\frac{c^{'}\delta_{sc}^{2}}{D^{2}}\right)^{a^{'}} \qquad b^{'}=\frac{(2c)^{-\frac{a}{2}}}{b}
\end{equation}

\end{document}